\documentclass[12pt,preprint]{aastex}
\usepackage{natbib}

\newcommand\teff{T_{\rm eff}}
\newcommand\fsed{f_{\rm sed}}

\shorttitle{Colors and Spectra of Directly Imaged Planets}
\shortauthors{Marley et al.}

\begin{document}

\title{Masses, Radii, and Cloud Properties of the HR 8799 Planets}

\author{Mark S. Marley}
\affil{NASA Ames Research Center, MS-245-3, Moffett Field, CA 94035; Mark.S.Marley@NASA.gov}

\author{Didier Saumon}
\affil{Los Alamos National Laboratory, Mail Stop F663, Los Alamos NM 87545; dsaumon@lanl.gov}

\author{Michael Cushing}
\affil{Department of Physics and Astronomy, The University of Toledo, 2801 West Bancroft Street, Toledo, OH 43606; michael.cushing@utoledo.edu}

\author{Andrew S.\ Ackerman}
\affil{NASA Goddard Institute for Space Studies, 2880 Broadway, New York, NY 10025; andrew.ackerman@nasa.gov}

\author{Jonathan J. Fortney}
\affil{Department of Astronomy and Astrophysics, University of California, Santa Cruz, CA 95064; jfortney@ucolick.org}

\and
\author{Richard Freedman}
\affil{SETI Institute \& NASA Ames Research Center, MS-245-3, Moffett Field, CA 94035,  
U.S.A.; freedman@darkstar.arc.nasa.gov}

\begin{abstract}
The near-infrared colors of the planets directly imaged around the A star HR 8799  are much redder than most field brown dwarfs of the same effective temperature.  Previous theoretical studies of these objects have concluded that the atmospheres of planets b, c, and d are unusually cloudy or have unusual cloud properties.  Some studies have also found that the inferred radii of some or all of the planets disagree with expectations of standard giant planet evolution models.  Here we compare the available data to the predictions of our own set of atmospheric and evolution models that have been extensively tested against observations of field L and T dwarfs, including the reddest L dwarfs.  We require mutually consistent choices for effective temperature, gravity, cloud properties, and planetary radius.  This procedure thus yields plausible values for the masses, effective temperatures, and cloud properties of all three planets.  We find that the cloud properties of the HR 8799 planets are  not unusual but rather follow previously recognized trends, including a gravity dependence on the temperature of the L to T spectral transition--some reasons for which we discuss.   We find the inferred mass of planet b is highly sensitive to whether or not we include the $H$ and $K$ band spectrum in our analysis.   Solutions for planets c and d are consistent with the generally accepted constraints on the age of the primary star and orbital dynamics.  We also confirm that, like in L and T dwarfs and solar system giant planets, non-equilibrium chemistry driven by atmospheric mixing is also important for these objects. Given the preponderance of data suggesting that the L to T spectral type transition is gravity dependent, we present an exploratory evolution calculation that accounts for this effect.  Finally we recompute the the bolometric luminosity of all three planets.

\end{abstract}


\section{INTRODUCTION}

Establishing the masses, radii, effective temperatures, and atmospheric composition of the planets orbiting the A star HR 8799 has been a challenge.  Of the four planets \citep{Mar08, Mar10} directly imaged orbiting the star HR 8799, broad photometric coverage (1 -- $5\,\rm \mu m$) is available for three planets, b, c, and d \citep{Mar08, Cur11}, and some spectral data is available for one planet, b \citep{Bar11a}.  Efforts to fit the available data with atmosphere and evolution models have produced mixed results.  In some cases the best-fitting models predict radii and ages that are at odds with other constraints, such as evolution models and the age of the system.  The purportedly unusual cloud properties of the planets have also received great attention.  Here we present an examination of the properties of HR 8799 b, c, and d using publicly available data as well as our own evolution and atmosphere models.   Our aim is to determine if a set of planet properties can be derived that simultaneously satisfy all observational and theoretical constraints and to ascertain the nature of atmospheric condensate layers in each planet.

We open below with a summary of the model parameters previously derived for these planets. In the remainder of this section we briefly review what is known about the atmospheric evolution of brown dwarfs and discuss the issues that have arisen to date in the study of the HR 8799 planets, particularly regarding the inferred cloud properties and planet radii.  In succeeding sections we explore the nature of clouds in low-mass objects more deeply and present model solutions for the masses, effective temperatures  $(T_{\rm eff}$), and cloud properties of the planets.  We find, as have all other previous studies, that clouds are present in the visible atmosphere of these planets at lower effective temperatures than in typical field brown dwarfs. In agreement with \citet{Bar11a} but unlike most other previous studies \citep[e.g.,][]{Bow10,Cur11,Mad11} we find that the clouds of the HR 8799 planets are similar to those found in field L dwarfs.  

\subsection{Masses and Radii of HR 8799 b,c, and d}
In the HR 8799 b, c, and d discovery paper, \citet{Mar08} derived the mass and effective temperature of each object in two ways.  In the first method they computed the luminosity of each object and compared that to theoretical cooling tracks for young giant planets given the constraint of their estimated age of the primary star.  In the second method they fit atmosphere models derived using the PHOENIX code \citep{Hau99} to the available six-band near-infrared photometry (1 to $2.5\,\rm \mu m$) to constrain 
$T_{\rm eff}$ and $\log g$, the two most important tunable parameters of atmosphere models.  Radii of each planet were derived by comparing the model emergent spectra with the observed photometry and known distance to the target.  Notably only models that included the effects of refractory silicate and iron clouds were consistent with the data.  However the radii estimated by this method were far smaller than expected for solar metallicity gas giant planets at such young ages. 

A number of followup studies presented new data new data and models in an attempt to better understand the planets.
\citet{Bar11a} fit a suite of models to the available photometry (but not the $M$ band \citep{Gal11} data) and $H$ and $K$ band spectra that they obtained for planet b.  By comparing the integrated flux from their best fitting model atmosphere to the estimated bolometric luminosity of the planet, they found a  small radius for the planet $R\sim0.75\,\rm R_J$. 
\citet{Gal11} also fit the Barman atmosphere models to the photometry, including new $M$ band data.  They found somewhat higher gravity solutions than \citet{Bar11a}  but also required  a small radius for planet b, approximately 70\%--or about one-third the volume--expected from planetary evolution models.  Such a large discrepancy is difficult to reconcile with our understanding of both giant planet evolution and the high pressure equation of state of hydrogen.  Instead the most straightforward interpretation is that the atmosphere models are not representative of the actual planetary atmosphere and Barman et al.\ suggest that higher metallicity models might provide a better fit and give more plausible radii.
 
Likewise \citet{Bow10} selected the model spectra (from among the models of \citet{Hub07,Bur06,All01}) which best fit the available photometry for HR 8799b.  Their best fitting spectra were quite warm, with $T_{\rm eff}$ from 1300 to 1700 K and thus they required even smaller radii ($\sim 0.4\,\rm R_J$) in order to meet the total luminosity constraint given the photometry available at that time.

In contrast \citet{Cur11} searched for the best fitting models while requiring that the planet radii either matched those predicted by a set of evolution models \citep{Bur97} or were allowed to vary.  They found that what they termed to be ``standard'' brown dwarf cloud models required unphysically small planet radii to fit the data.  However their ``thick cloud'' models could fit the data shortward of $3\,\rm \mu m$ by employing planetary radii that were within about 10\% of the usual evolution model prediction.  As we note below, however, the ``standard'' cloud model has itself not been demonstrated to fit cloudy, late L-type dwarfs; thus this exercise does not necessarily imply the  planets' clouds are ``non-standard''.  Nevertheless they were able to fit much of the photometry with planetary radii consistent with evolution model predictions.

Finally \citet{Mad11} explored  a set of models similar to those studied by Currie et al.\ with yet another cloud model but without the radius constraint.  Their best fits are very similar to those of Currie et al.\ but with somewhat lower $T_{\rm eff}$.

The characteristics of the planets as derived in the 2011 publications are summarized in Table 1.  Not all authors report every parameter so some radii and ages are left blank.  Note the diverse set of masses, radii, and effective temperatures derived by the various studies.  Despite the variety some trends are clear: planet b consistently is found to have the lowest mass and effective temperature and its derived radius is almost always at odds with the expectation of evolution and interior models. 

We note that at very young ages the model radii of giant planets depends on the initial conditions of the evolutionary calculation \citep{Ste82,Bar02,Mar07a,Spi12}.  However at ages younger than several hundred million years the  planetary radius is expected to be no smaller than about 1.1 times that of Jupiter regardless of the formation mechanism.  Hence radii derived by \citet{Bar11a} and \citet{Gal11} are not consistent with evolutionary calculations, regardless of the initial boundary conditions.  Indeed the equation of state for gas giant planets, even ones  enriched in heavy elements, preclude such radii.

\subsection{Clouds}
\subsubsection{Brown Dwarfs}
As a brown dwarf ages it radiates and cools.  When it is warm, refractory condensates, including iron and various silicates, form clouds in the visible atmosphere.  Over time the clouds become progressively thicker and more opaque, leading to ever redder near-infrared colors.  As the dwarf cools the cloud decks are found at higher pressures, deeper in the atmosphere. Eventually the clouds disappear from the photosphere.  Indeed the first two brown dwarfs to be discovered, GD 165B \citep{Beck88} and Gl 229B \citep{Nak95}, were ultimately understood to represent these two different end cases: the cloudy L and the clear T dwarfs (see \citet{Kir05} for a review). Understanding the behavior of clouds in substellar atmospheres and how it might vary with gravity has become one of the central thrusts of brown dwarf science.  
   
   The earliest models for these objects assumed that the condensates were uniformly distributed vertically throughout the atmosphere \citep[e.g.,][]{Chab00}.  Later, more sophisticated approaches attempted to model the formation of discrete cloud layers that would result from the gravitational settling of grains.  

With falling effective temperature, $T_{\rm eff}$, the bases of the iron and silicate cloud decks are found progressively deeper in the atmosphere.  Because of grain settling the overlying atmosphere well above the cloud deck loses grain opacity and becomes progressively cooler. Thus over time more of the visible atmosphere becomes grain free and cooler.  Cooler temperatures favor  $\rm CH_4$ over CO.  The removal of the opacity floor that the clouds provided at higher $T_{\rm eff}$ also allows flux in the water window regions to escape from deeper in the atmosphere.  This leads to a brightening in the $J$ band and a blueward color shift in the near-infrared.  
In field brown dwarfs this color change begins around effective temperature $T_{\rm eff}\sim 1200$ to $1400\, \rm K$ and is complete over a strikingly small effective temperature range of only 100 to 200$\,$K (see \citet{Kir05} for a review).
This experience led to the presumption that all objects with effective temperatures below about 1100 K would have blue near-infrared colors, like the field brown dwarfs.  

\subsubsection{HR 8799 b, c, and d}
The early directly imaged low mass companions  confounded these expectations from the brown dwarf experience.  The companion  2MASSW J1207334-393254 b (hereafter 2M1207 b) has red infrared colors despite its low luminosity and apparently cool $T_{\rm eff}$ \citep{Cha04} .  Likewise the HR 8799 planets  have colors reminiscent of hot, cloudy L dwarfs but their bolometric luminosities coupled with radii from planetary structure calculations imply $T_{\rm eff}\sim 1000\,\rm K$ or lower \citep{Mar08,Mar10}.  

The red colors, particularly of the HR 8799 planets, spawned a storm of studies investigating the atmospheric structure of the planets.  Essentially all of these papers concluded that the planets could be best explained by invoking thick cloud decks.  Since this ran counter to expectation, these clouds were deemed ``radically enhanced'' when compared to ``standard'' models \citep{Bow10}.   Likewise  \citet{Cur11} compared their data to the \citet{Bur06} model sequence and concluded (their \S5) that the HR 8799 planets have much thicker clouds than ``...standard L/T dwarf atmosphere models.''   \citet{Mad11} state that their fiducial models ``...have been shown to provide good fits to observations of L and T dwarfs \citep{Bur06}''.  They then find that much cloudier models are required to fit the imaged exoplanets and thus conclude that the cloud properties must be highly discrepant from those of the field L dwarfs.  

Such conclusions, however, seem to overlook that the study of L dwarf atmospheres is still in its youth.  Cloudy atmospheres of all kinds are challenging to model and the L dwarfs have proven to be no exception.  Thus whether or not the HR 8799 planets have unusual clouds depends on the point of reference.  Indeed while most published models of  brown dwarfs are able to reproduce the spectra of cloudy, early L-type dwarfs and cloudless T dwarfs, the latest, reddest---and presumably cloudiest---L dwarfs have been a challenge.   The points of comparison for the work of
\citet{Cur11} and \citet{Mad11} were the models described in \citet{Bur06}.  When compared to the red-optical and near-infrared photometry of L and T dwarfs, those models did not reproduce the colors of the latest L dwarfs as the models are too blue (see figure 17 of \citet{Bur06}) implying that they lacked sufficient clouds. \citet{Bur06}   also presented comparisons of their models to L dwarf spectra; however the comparisons are only to an L1 and an L5 dwarf.  There are no comparisons to very cloudy late L dwarf spectra in the paper so the fidelity of their model under such conditions cannot be judged.  For these reasons a comparison of the cloudy HR 8799 planets to the  ``standard'' L dwarf models, such as presented by \citet{Mad11} and \citet{Cur11},  does not address the question whether the HR 8799 planets are really all that different from the cloudiest late L dwarfs since those models have apparently do not reproduce the colors of the latest L dwarfs.

At least one set of atmosphere and evolution models is available that has been compared against the near- to mid-infrared spectra and colors of latest L dwarfs.  In \citet{Cus08} and \citet{Ste09} we compared our group's models to observed far-red to mid-infrared spectra of L and T dwarfs, including L dwarfs with IR spectral types as late as L9 (with 7 objects in the range L7 to L9.5).  We found that the models with our usual cloud prescription fit the spectra of L dwarfs of all spectral classes (including the latest field dwarfs) well, but not perfectly.  In \citet{Sau08} we also presented a model of brown dwarf evolution that well reproduced the usual near-infrared color magnitude diagrams of L and T dwarfs, including the reddest L dwarfs. Here we apply our set of cloudy evolution models to the HR 8799 planet observations in an attempt to better understand these objects.

\subsection{Chemical Mixing}
Shortly after the discovery of  Gl 229B, \citet{Feg96}
predicted that---as in Jupiter---vertical mixing might cause CO to be 
overabundant compared to $\rm CH_4$ in chemical equilibrium in this object. This was promptly confirmed by the detection of CO 
absorption at $4.6\,\rm \mu m$ by \citet{Nol97} and \citet{Opp98}.  The overabundance is caused by
the slow conversions of CO to $\rm CH_4$ relative to the mixing time scale.

An obvious mechanism for vertical mixing in an atmosphere is convection.  
Brown dwarf atmospheres are convective at depth where the mixing time scale is 
short (minutes).  The overlying radiative zone is usually considered quiescent but a 
variety of processes can cause vertical mixing, albeit on much longer time scales.
 Since the conversion time scales for $\rm CO \rightarrow CH_4$ and $\rm N_2 \rightarrow NH_3$ range from 
seconds (at $T\sim 3000\,\rm K$) to many Hubble times (for $T< 1000\,\rm K$), even very slow mixing in 
the radiative zone can drive the chemistry of carbon and nitrogen out of equilibrium. 
From this basic consideration, it appears that departures from equilibrium are inevitable 
in the atmospheres of cool brown dwarfs and indeed the phenomenon is well established \citep[e.g.,][]{Sau00, Geb01, Hub07,Geb09, Mai07, Sau06, Ste09}.

With falling gravity the point at which chemical reactions are quenched occurs deeper in the atmosphere, where the higher temperature result in a greater atmospheric abundance of CO \citep{Hub07,Bar11a}.  At exoplanet gravities, mixing can even produce CO/$\rm CH_4$ ratios in excess of 1 \citep{Bar11a}.  Thus a complete giant planet exoplanet atmosphere model must account for such departures from chemical equilibrium as well.

\section{Gravity, Refractory Clouds and the L/T Transition}

\subsection{Nature of the Transition}

Two main  causes of the loss of cloud opacity at the L to T transition have been suggested.  In one view the atmospheric dynamical state changes, resulting in larger particle sizes that fall out of the atmosphere more rapidly, leading to a sudden clearing or collapse of the cloud \citep{Kna04, Tsu03, Tsu04}.  This view is supported by fits of spectra to model spectra \citep{Sau08}  computed with the \citet{Ack01} cloud model.  In that formalism, a tunable parameter, $f_{\rm sed}$ controls cloud particle sizes and optical depth.  Larger $f_{\rm sed}$ yields larger particles along with physically and optically thinner clouds.  \citet{Cus08} and \citet{Ste09} have demonstrated that progressively later  dwarfs (L9 to T4) can be fit by increasing $f_{\rm sed}$ across the transition at a nearly fixed effective temperature.  A variation on this hypothesis is that a cloud particle size change is responsible for the transition \citep{Bur06}.

The second view is inspired by thermal infrared images of the atmospheres of Jupiter and Saturn at $\sim 5\,\mu$m \citep[e.g.][]{Wes69,Wes74,Ort96,Bai05}.  Gaseous opacity is low at this wavelength and the clouds stand out as dark, mottled features against a bright background of flux emitted from deeper, warmer levels in the atmosphere.  Such images of both Jupiter and Saturn clearly show that the global cloud decks are not homogenous, but rather are quite patchy.  
\citet{Ack01}, \citet{Bur02}, and \citet{Marl10} have suggested that the arrival of holes in brown dwarf clouds, perhaps due to the clouds passing through a dynamical boundary in the atmosphere, might also be responsible for the L to T transition.    This view is supported by the discovery of L-T transition dwarfs that vary in brightness with time with relatively large near-infrared amplitudes \citep{Art09,Rad11}.  Indeed Radigan (in prep) has found in a survey of about 60 L and T type brown dwarfs that the most variable dwarfs are the early T's, which are  in the midst of the $J-K$ color change.

In order to match observations, modern thermal evolution models for the cooling of brown dwarfs have to impose some arbitrary mechanism, such as varying sedimentation efficiency or the imposition of cloud holes, by which the thick clouds in the late L dwarfs dissipate.  A uniform cloud layer that simply sinks with falling $T_{\rm eff}$ as the atmosphere cools turns to the blue much more slowly than is observed.  Application of such a transition mechanism to reliably reproduce the colors and spectra of late L and early T dwarfs (e.g., near-infrared color-magnitude diagrams) led to the expectation that the normal behavior for cooling brown dwarfs--or extrasolar giant planets--is to turn to the blue at around 1300 K.  

However there have been indications that such a narrative is too simplistic and that gravity plays a role as well.  Two brown dwarf companions to young main sequence stars were found to have unexpectedly cool effective temperatures for their L-T transition  spectral types
by \citet{Met06} and \citet{Luh07}.  The analysis of Luhman et al.\ of the T dwarf HN Peg B was further supported by additional modeling presented in \citet{Leg08}.  \citet{Dup09} presented evidence of a gravity dependent transition $T_{\rm eff}$ on the basis of a dynamical mass determination of an $\rm M8 + L7$ binary.  
\citet{Ste09} fit the model spectra of \cite{Mar02} to the 1 -- $15\,\rm \mu m$ spectra of L and T dwarfs and found that L dwarf cloud clearing (as characterized by large $f_{\rm sed}$) occurs at $T_{\rm eff} \sim 1300\,\rm K$ for $\log g = 5.0$ and at $\sim 1100\,\rm K$ for $\log g = 4.5$, although the sample size was 
admittedly small (Figure 1).  Nevertheless such an association implies a cooler transition temperature at even lower gravity.

\subsection{Clouds at Low Gravity}

Even if directly imaged planets are not considered, there is already considerable evidence that the cloud clearing associated with the L to T transition occurs at lower effective temperatures in lower gravity objects than in high gravity ones.  To understand what underlies this trend it is necessary to consider three separate questions. First, where does the optically-thick portion of the cloud lie in the atmosphere relative to the photosphere, as a function of gravity?  An optically-thick cloud lying well below the photosphere will be essentially invisible whereas the same cloud lying higher in the atmosphere would be easily detected.  Second, how does the total optical depth of the cloud vary with gravity?  This is a complex problem involving the pressure of the cloud base and the particle size distribution.  Third, how does the mechanism by which clouds dissipate vary with gravity?  For example, do holes form at a different effective temperature in different gravity objects?  In this section we consider only the first two questions and defer the third question to Section 5.6.

To address the first question we need to understand how atmospheric temperature $T$ varies with pressure $P$ as a function of gravity.  For a fixed effective temperature, a lower gravity atmosphere is warmer at a fixed pressure level than a higher gravity one.  This is because more atmospheric mass--and thus greater opacity--overlies a given pressure level at lower gravity.   Figure 2 provides an example using our model profiles.  Since at equilibrium condensation begins at the intersection of the vapor pressure and thermal profiles, the cloud base occurs at lower pressure (higher in the atmosphere) in a low gravity object than a high gravity one.

As objects cool with time (at essentially fixed gravity) clouds will persist at lower pressure and remain visible to cooler effective temperatures in lower gravity objects than higher gravity ones.  For example in Figure 2 the lowest gravity model shown at $T_{\rm eff}=900\,\rm K$ is hotter at all pressures greater than a few hundred millibar than a higher gravity $T_{\rm eff}=1300\,\rm K$ object. As explained below this degeneracy between cooler low gravity and warmer high gravity temperature profiles lies at the heart of the problem of simultaneously distinguishing gravity and effective temperature with a limited photometric dataset.

Addressing the second question requires us to understand how the cloud column optical depth varies with gravity.  This depends both on 
the amount of condensible material  in the atmosphere available to form clouds and on the cloud particle size.  From basic scaling laws and mass balance \citet{Mar00} derived an expression for the wavelength-dependent total column optical depth $\tau_\lambda$ of a cloud in a hydrostatic atmosphere
$$\tau_\lambda = 75 \epsilon Q_\lambda(r_{\rm eff})\varphi{\biggl({P_{cl}\over {1\,\rm bar}}\biggr)}  {\biggl({10^5\,{\rm cm\,s^{-2}}\over {g}}\biggr)} {\biggl({1\,{\rm \mu m}\over {r_{\rm eff}}}\biggr)} {\biggl({1.0\,{\rm g\,cm^{-3}}\over {\rho_c}}\biggr)}.  \eqno(1)    $$
Here $P_{cl}$, $r_{\rm eff}$ and $\rho_c$ refer to the pressure at the cloud base and the condensate effective (area-weighted) radius\footnote{Marley (2000) employed the mean particle size $r_c$ rather than the more rigorous area-weighted size.} and density (see also Eq. 18 of \citet{Ack01}).  $\varphi$ is the product of the condensing species number mixing ratio and the ratio of the mean molecular weight of the condensate to that of the atmosphere. The expression assumes that some fraction $\epsilon$ of the available mass above the cloud base forms particles with extinction cross section $Q_\lambda$  (which can be computed through Mie theory) .   \citet{Ack01} also estimate the column optical depth of a cloud with a similar result.  Generalizing their Eq. 16,
$$\tau_\lambda \propto {P_{cl} \over{g r_{\rm eff} (1+f_{\rm sed})} }. \eqno(2)    $$

Both Equations (1) and (2) hold that all else being equal--including particle sizes--we expect $\tau \propto P_{cl}/g$, just because the column mass above a fixed pressure level is greater at low gravity and there is more material to condense.  Any cloud model which self-consistently computes the column mass of condensed material should reproduce this result.  As shown above, however, the cloud base is at lower pressure in lower gravity objects, roughly $P_{cl} \propto g$, thus leading to the expectation that the cloud $\tau$ would be approximately constant with changing gravity.  This is not exactly true since there is a slope to the vapor pressure equilibrium curve and thus the actual variation is somewhat weaker, but  the effects of gravity and the cloud base pressure alone do not strongly influence cloud column optical depth.

The second component affecting the column cloud opacity is particle size.  While a cloud model is required for rigorous particle size computation, we can examine the scaling of size with gravity. At lower gravity particle fall speeds are reduced, which reduces the downward mass flux carried by condensates of a given size $r$.  Since fall speed is proportional to $r^2$ in the Stokes limit (the viscous regime at low Reynolds numbers) while the mass is proportional to $r^3$, the flux scales with $r^5$, a slight increase in particle size can produce the same mass balance in the atmosphere at lower gravity, and thus $r$ is expected to increase relatively slowly with decreasing $g$.  At large Reynolds number the dependence on fall speed is weaker than $r^2$ and the equivalent result is found. Indeed recasting the \citet{Ack01} model equations suggests $r\propto (f_{\rm sed} / g)^{1/2}$, although the actual dependence is more complex as it depends upon an integral over the size distribution. Tests with the complete cloud model coupled to our atmosphere code predict about a factor of 4 increase in cloud particle radius (25 to $100\,\rm \mu m$) as gravity decreases by an order of magnitude from 300 to $30\,\rm m\,s^{-2}$, a slightly faster increase than $\sqrt{g}$.  A roughly  $r\propto g^{-1/2}$ relationship is also seen in the cloud model of \citet{Coo03} (see their Figures 2, 3, and 4). Returning to Eq.\  (1) and combining with the scaling discussed above thus suggests that all else being equal we expect cloud $\tau \propto \sqrt{g}$.

Figure 3 illustrates all of these effects in model cloud profiles calculated for three atmosphere models with varying $g$ and $T_{\rm eff}$.  The atmospheric gravity spans two orders of magnitude while the effective temperature varies from 1200 to 1000 K from the warmest to coolest object.  As expected the cloud particle size indeed varies inversely with gravity($r \sim g^{-1/2}$) while the cloud base pressure decreases with decreasing gravity.  The choice in the plot of a cooler $T_{\rm eff}$ for the lowest gravity object counteracts what would otherwise be an even greater difference in the cloud base pressure.  The net result is that the total column optical depth for the silicate cloud in all three objects is very similar, $\tau\sim 10$.  {\em Thus a cooler, low gravity object has a cloud with a column optical depth that is almost indistinguishable from that of a warmer, more massive object.}

The thicker portion of the lines denoting cloud column optical depth signify the regions in the atmosphere where the brightness temperatures between  $\lambda = 1$ and $6\,\rm \mu m$ are equal to the local temperature.  In other words the thick line represents the near-infrared photosphere.  In all three cases there is substantial cloud optical depth ($\tau_{\lambda}>0.1$) in the deeper atmospheric regions from which flux emerges in the near-infrared.  As a result clouds play comparable roles in all three objects despite the two order of magnitude difference in gravity and the 200 K temperature difference.  We thus conclude that the net effect of all of these terms is to produce clouds in lower gravity objects with optical depths and physical locations relative to the photosphere comparable to clouds in objects with higher gravity and higher effective temperature. 
 
\section{Modeling Approach}
To model the atmospheres and evolution of exoplanets we apply our usual modeling approach which we briefly summarize in this section.  We stress that the fidelity of model fits in previous applications of our method to both cloudy and clear atmosphere brown dwarfs \citep{Mar96, Mar02, Bur97, Roe04, Sau06, Sau07, Leg07a, Leg07b, Mai07, Bla07, Cus08, Geb09, Ste09} validates our overall approach and provides a basis of comparison to the directly imaged planet analysis.  In addition  to brown dwarfs the model has been applied to Uranus \citep{Mar99} and Titan \citep{Mck89} as well.
   
\subsection{Atmosphere and Cloud Models}

The atmospheric structure calculation is described in \citet{Mck89, Mar96, Bur97, Mar99, Mar02,  Sau08}.  Briefly
we solve for a radiative-convective equilibrium thermal profile that carries thermal flux given by $\sigma T_{\rm eff}^4$ given
a specified gravity and atmospheric composition. 
The thermal radiative transfer follows the source function technique of \citet{Too89}
allowing inclusion of arbitrary Mie scattering particles in 
the opacity of each layer. Our opacity database includes all important absorbers and is described in \citet{Fre08}. 

There are, however, two particularly important updates to our opacity database since \citet{Fre08}.  First we use a new molecular line list for ammonia \citep{Yur11}.   Secondly we have updated our previous treatment of pressure-induced opacity arising from collisions of $\rm H_2$ molecules with $\rm H_2$ and He.  This new opacity is discussed in \citet{Fro10} and the impact on our model spectra and photometry in general is discussed in \citet{Sau12}.

The abundances of molecular, atomic, and ionic species are computed for chemical equilibrium
as a function of temperature, pressure, and metallicity following \citet{Feg94, Feg96, Lod99, Lod02, Lod03, Lod06}
assuming the elemental abundances of \citet{Lod03}.  In this paper we explore only solar composition models.  

For cloud modeling we employ the approach of  \citet{Ack01} which parameterizes the importance of sedimentation relative to upwards mixing of cloud particles through an efficiency factor, $f_{\rm sed}$.  Large values of $f_{\rm sed}$ correspond to rapid particle growth 
and large mean particle sizes. Under such conditions condensates quickly 
fall out of the atmosphere, leading to physically and optically 
thinner clouds. In the case of small $f_{\rm sed}$  particles grow more slowly resulting in a larger 
atmospheric condensate load and thicker clouds. Both our 
cloud model and chemical equilibrium calculations are fully coupled with the radiative transfer and the 
$(P, T )$ structure of the model during the calculation of a model so 
that they are fully consistent when convergence is obtained. 

We note in passing that the cloud models employed in previous studies of the HR 8799 planets have been {\it ad hoc}, as straightforwardly
discussed in those papers.  Particle sizes, cloud heights, and other cloud properties are fixed at given values while gravity, $T_{\rm eff}$, and other model parameters are varied.  The methodology used here is distinct since in each case we compute a consistent set of cloud properties given a specific modeling approach, the Ackerman \& Marley cloud.

The coupled cloud and atmosphere models have been widely compared to spectra and photometry of L and T dwarfs in the publications cited in the introduction to this section.  We emphasize in particular that \citet{Cus08} and \citet{Ste09}  show generally good fits between our model 
spectra and observations of cloudy L dwarfs. The near-infrared colors of brown dwarfs are quite sensitive to the choice of $f_{\rm sed}$, a 
point we will return to in Section 5.4.

\subsection{Evolution Model}
Our evolution model is described in \citet{Sau08}. In fitting the HR 8799 data,
we use the sequence computed with a surface boundary condition extracted from 
our cloudy model atmospheres with $\fsed=2$. As we will see below, our best fits
show that all three planets are cloudy with $\fsed=2$, which justifies this choice
of evolution {\it a posteriori}. As the three planets appear to have
significant cloud decks (as will be confirmed below), it is not necessary to 
use evolution sequences that take into account the transition explicitly 
in this comparison with models.  Nevertheless, we will explore the effects of a gravity-dependent
transition between cloudy and cloudless atmospheres in Section 5.4 as this is a topic of
growing interest.

The \citet{Sau08} models were computed with what has come to be known as a traditional or hot-start initial condition.  As discussed in \citet{Bar02}, \citet{Mar07a} and \citet{Spi12} however, the computed radii of young giant planets at ages of 100 Myr and less is highly dependent on the details of the assumed initial condition.  Even assuming very cold initial conditions, however, computed planetary radii never fall below $1\,\rm R_J$ at ages of less than 1 Gyr.    Rather than carrying out the model fitting for an uncertain set of assumed cold initial conditions, we choose here to employ the traditional hot-start boundary conditions for the evolution modeling.  In this way we avoid unphysical very small radii ($R < 1\,\rm R_J$) while adding an  additional constraint to the modeling.

\section{Application to HR 8799 Planets} 

\subsection{Constraints on the HR 8799 System Properties}

A number of the properties of the HR 8799 system as a whole help to constrain the properties of the individual planets.  Of foremost importance of course is the age of the primary star since older ages require greater planetary masses to provide a fixed observed luminosity.  The massive dust disk found outside of the orbit of the most distant planet, HR 8799 b, constrains the mass of that planet since a very massive planet would disrupt the disk.  Finally dynamical models of the planetary orbits circumscribe the parameter
space of orbits and masses that are stable over the age of the system.  All of these topics have been discussed extensively in the literature so here we briefly summarize the current state of affairs.  A more thorough review can be found in \citet{Sud12}.

Since the discovery of the first three planets, the age of HR 8799 has been debated.  As summarized initially by the discoverers, most indicators suggest a young age of 30 to 60 Myr \citep{Mar08}. However the typical age metrics are somewhat more in doubt than usual because HR 8799 is a $\lambda$ Boo-type star with an unusual atmospheric and uncertain internal composition. \citet{Moy10} review the various estimates of the age of the star prior to 2010 and argue that most of the applied metrics, including color and position on the HR diagram,  are not definitive.  Most recently \citet{Zuc11} conclude that the Galactic space motion of HR 8799 is very similar to that of the 30 Myr old Columba association and suggest that it is a member of that group. They also argue that the $B-V$ color of HR 8799 in comparison to Pleiades A stars also supports a young age, although the unusual composition hampers such an argument.   Perhaps the fairest summary of the situation to date would be that most traditional indicators support a young age for the primary, but that no single indicator is entirely definitive on its own.  

One indicator that the age could be much greater than usually assumed is discussed by \citet{Moy10}.  Those authors use the $\gamma$ Doradus g-mode pulsations of the star to place an independent constraint on the stellar age.  Their analysis is dependent upon the rotation rate of the star and consequently the unknown inclination angle and thus is also uncertain.  Nevertheless they find model solutions that match the observed properties of the star in which the stellar age can plausibly be in excess of 100 Myr and in some cases as large as 1 Gyr or more.  They state that their analysis is most uncertain for inclination angles in the range of 18 to $36^\circ$, which corresponds to the likely inclination  supported by observations of the surrounding dust belt (see below).  Thus stellar seismology provides an intriguing, but likewise still uncertain constraint.  

The dust disk encircling the orbits of the HR 8799 planets can in principle provide several useful constraints on the planetary masses and orbits.  First the inclination of the disk affects the computed orbital stability of the companions \citep{Fab10} if we assume the disk is coplanar with the planetary orbits.  If the rotation axis of the star is perpendicular to the disk, the inclination also has a bearing on the stellar age since the seismological analysis in turn depends upon its inclination to our line of sight \citep{Moy10}.  \citet{Hug11}  discuss a variety of lines of evidence that bear on the inclination, $i$, of the HR 8799 dust disk.  While they conclude that inclinations near $20^\circ$ are most likely, the available data cannot rule out a face-on ($i=0^\circ$) configuration.  Finally an additional important constraint on the mass of HR 8799 b could be obtained if it is responsible for truncating the inner edge of the dust disk.  An inner edge at 150 AU is consistent with available data \citep{Su09} and this permits HR 8799 b to have a mass as large as $20\,\rm M_J$ \citep{Fab10}.  It is worth noting, however, that this limit depends upon the model-dependent inner edge of the disk and the dynamical simulations.  

Finally dynamical simulations of the planetary orbits constrained by the available astrometric data can provide planetary mass limits. In the most thorough study to date \citet{Fab10} found that if planets c and d were in a 2:1 mean-motion resonance their masses could be no larger than about $10\,\rm M_J$.  However if there were a double resonance in which c, d, and b participated in a ``double 2:1'' or 1:2:4 resonance (originally identified by \cite{Goz09}) then masses  as large as $20\,\rm M_J$ are permitted and such systems are stable for 160 Myr \citep{Fab10}.  Such a resonance was found to be consistent with the limited baseline of astrometric data. HR 8799 b,c, and d have also been identified in an archived HST image taken in 1998 \citep{Laf09, Sou11}. These data continue to allow the possibility of the 1:2:4 mean motion resonance, a solution which implies a moderate inclination ($i=28^\circ$) for the system.  New dynamical models that include both this new astrometric data and the innermost e planet are now required to fully evaluate the system's stability.  \citet{Sud12} studied such a system with masses of 7, 10, 10, and $10\,M_{\rm J}$.  They generally found system lifetimes shorter than 50 Myr for such large masses but at least one system was found to be stable for almost 160 Myr.  

Taken as a whole the age of the system and the available astrometric data and dynamical models are consistent with a relatively young age (30 to 60 Myr) and low masses for the planets (below $10\,\rm M_J$).  However the possibility of an older system age, as allowed by the asteroseismology, and higher planet masses, as permitted if the planets are in resonance and by the dust disk dynamics, cannot be fully ruled out.  Given this background we now consider the planetary atmosphere models.

\subsection{Data Sources}

The available photometric data for each planet is summarized in Table 2 and shown on  Figures \ref{planet_b}--\ref{planet_d}.  In addition for planet b we employ $H$ and $K$ band spectra as tabulated in \citet{Bar11a}.  We do not include the narrow band photometry of \citet{Bar11a} since this dataset has been superseded by the spectroscopy.  We also do not include very recent 3.3-$\rm \mu m$ photometry from \citet{Ske12} which became available after the submission of this manuscript although we do plot the point in Figures \ref{planet_b}--\ref{planet_d}. Below we summarize the sources of the photometry used in the fitting.  With the exception of the Subaru z-band which sits in an atmospheric window, we included 
      an atmospheric transmission curve when computing the synthetic magnitudes of the model 
      spectra.  The transmission curve was generated with ATRAN (Lord 1992) at an airmass of 1 
      with a precipitable water vapor content of $2\,\rm mm$.

\subsubsection{Subaru-$z$ band}

  The Subaru-$z$-band photometry is from \citet{Cur11} and   was obtained with the Infrared Camera and Spectrograph (IRCS; \citet{Tok98}) on the Subaru Telescope.  The filter   profile was kindly provided by Tae-Soo Pyo.  No atmospheric absorption was included
  because the filter sits in a window that is nearly perfectly   transparent.

\subsubsection{$J$ band} 

  The $J$ band data were taken from \citet{Mar08} and \citet{Cur11}.  The former
  observations were done with the Near-Infrared Camera (NIRC2) on Keck
  II which uses a Mauna Kea Observatories Near-Infrared (MKO-NIR) $J$ band filter.  We used the filter
  transmission profile from \citet{Tok02}.  The latter
  observations were obtained with the Infrared Camera and Spectrograph
  (IRCS; \citet{Tok98}) on the Subaru Telescope   which also uses a MKO-NIR $J$ band filter.

\subsubsection{$H$  and $Ks$ bands} 

  The $H$-band and $K_s$-band data were taken from \citet{Mar08}.  The observations were done with the Near-Infrared Camera
  (NIRC2) on Keck II which uses MKO-NIR  filters.  We used the
  filter transmission profile from \citet{Tok02}.

\subsubsection{[3.3] band}

  The [3.3]-band data was taken from \citep{Cur11}.  The observations were done with the Clio camera at the MMT
  Telescope \citep{Fre04, Siv06}.  The filter is non standard and has a central wavelength of
  $3.3\,\rm \mu m$, and half-power points of 3.10 and  $3.5\,\rm \mu m$.  The filter
  transmission profile was provided by Phil Hinz.

\subsubsection{$L^\prime$ band}

  The $L^\prime$-band data was taken from \citet{Cur11}.
  The filter is the $L^\prime$ filter in the MKO-NIR system so we used the
  filter transmission profile from \citet{Tok02}.

\subsubsection{$M^\prime$-band}

  The $M$-band photometry of \citet{Gal11} was
  obtained using the Near-Infrared Camera (NIRC2) on Keck II. This filter profile is the same as the $M^\prime$ band of the MKO-NIR system.  We
  therefore used the filter transmission profile from \citet{Tok02}.

\subsection{Fitting Method}

In order to determine the atmospheric properties of the HR 8799 planets,
we compared the observed photometry to synthetic spectra generated from our
model atmospheres.  We used a grid of solar metallicity models with the
following parameters: $T_{\rm eff}=600$--$1300\,\rm K$ in steps of 50 K, $\log g =
3.5$--5.5 in steps of 0.25 dex, $f_{\rm sed}=1, 2$, and eddy mixing coefficient $K_{\rm zz}=0$, $10^4\,\rm cm ^2\, s^{-1}$.  
We identify the best fitting model and estimate the
atmospheric parameters of the planets following the technique described
in Cushing et al. (2012, in prep).  In brief, we use Bayes' theorem to
derive the joint posterior probability distribution of the atmospheric
parameters given the data $P(T_{\rm eff},\log g,f_{\rm sed},K_{\rm zz} |\mathbf{f})$, where
$\mathbf{f}$ represents a vector of the flux density values (or upper
limits) in each of the bandpasses.   Since the posterior distribution is only known to within a multiplicative constant, the practical outcome is a list of models ranked by their relative probabilities.

Estimates and uncertainties for each of the atmospheric parameters can
also be derived by first marginalizing over the other parameters and
then computing the mean and standard deviation of the resulting
distribution.  For example, the posterior distribution of $T_{\rm eff}$ is
given by,

$$P(T_{\rm eff}|\mathbf{f}) = \int P(T_{\rm eff},\log g,f_{\rm sed},K_{\rm zz}|\mathbf{f}) \,\, d\,\log g \,\,d f_{\rm sed} \,\, d\,K_{\rm zz}$$

\noindent
Since $(T_{\rm eff},\log g)$ values can be mapped directly to $(M, R, L_{\rm bol})$ values
using evolutionary models, we can also construct marginalized
distribution for $M$, $R$, and $L_{\rm bol}$.  Figure \ref{hist} shows the resulting distribution of
$T_{\rm eff}$, $\log g$, $M$, and $L_{\rm bol}$ for each planet and indicates the formal solution for
these parameters and and their associated uncertainties.

Finally note that we chose to use a Bayesian formalism rather than the more common approach of minimizing $\chi^2$ because 1) we can marginalize over  model parameters such as the distance and radii of the brown dwarfs, and 2) we can  incorporate upper limits using the formalism described in \citet{Iso86}.

\subsection{Results of  Model Fitting}
In this section we discuss the individual best fits to each planet.  Figures \ref{planet_b} -- \ref{planet_d} display the model fits to the observed spectra and photometry.
Each panel of Figure \ref{contour} shows contours, denoting integrated probabilities of 68, 95, and 99\%, in the $\log g - T_{\rm eff}$ plane. In these figures evolution tracks for planets and brown dwarfs of various masses are shown.  The objects evolve from right to left across the figures as they cool over time.  Isochrones for a few ages are shown;  the kinks arise from deuterium burning.  In some cases at a fixed age a given $T_{\rm eff}$ can correspond to three different possible masses (e.g., a 1150 K object  at 160 Myr).  Also shown are contours of constant $L_{\rm bol}$.  Note that the isochrones are derived from the conventional  hot-start  giant planet evolution calculation.  A different choice of initial conditions would result in different isochrones. 

The best fitting parameters are also shown in Figure \ref{hist} as histograms of probability distribution for $T_{\rm eff}$,  $\log g$,  $M$ and $L$. For $\log g$ and $T_{\rm eff}$ the histograms are projections of the contours shown in Figure \ref{contour} onto these two orthogonal axes.  The mean of the fit and the size of the standard deviation is indicated in each panel and also illustrated by the solid and dashed vertical lines. The third and fourth columns of Figure \ref{hist} depict the same information but for the mass and luminosity corresponding to each $(T_{\rm eff}, \log g)$ pair, as computed by the evolution model.

We discuss each set of fits for each planet in turn below.

\subsubsection{HR 8799b}
HR 8799b is the only one of the three planets considered here for which there is spectroscopic data and our results are sensitive to whether or not this data is included in our fit.  Contours which show the locus of the best fitting models for the photometry are shown in the left-hand panel of Figure \ref{contour}.  When only the photometric data is fit high masses around $\sim 26\,\rm M_J$ are favored.  The photometry-only fit finds $T_{\rm eff} = 1000\,\rm K$ and $f_{\rm sed}=2$ while a fit to both the spectroscopy and the photometry results in $T_{\rm eff} = 750\,\rm K$ and $f_{\rm sed}=2$ with a mass of $\sim 3\,\rm M_J$.  We reject the low temperature fit for several reasons: the solution lies at the edge of our model grid, such a planet would be very young, and  such a cold effective temperature is not consistent with the bolometric luminosity of planet b (see \S 5.2). These models are illustrated in the top two panels of Figure \ref{planet_b}.  

To isolate the effect of the spectroscopy of \citet{Bar11a} on the preferred fit, we relaxed the radius and distance constraint on the fitting and found the model that best reproduces the shape of the spectra.  Somewhat surprisingly this is a cold, very low gravity and very cloudy model ($T_{\rm eff} = 600\,\rm K$, $\log g=3.5$  and $f_{\rm sed}=1$).  With a standard radius such a model is again too young and faint and also lies at the edge of the model grid.

The reason the derived gravity depends so strongly on the $H$ and $K$ spectra is that the shape of the emergent flux--and not just the total flux in a given band--contains information about the gravity.  In particular a ``triangular'' $H$ band shape serves as an indicator of low gravity (see \citet{Ric11} and \citet{Bar11a}).  This shape results from the interplay of a continuum opacity source--either cloud opacity (in a cloudy atmosphere) or the collision-induced opacity of
molecular hydrogen (when cloud opacity is unimportant)--and a sawtooth-shaped water opacity (discussions in the literature generally only highlight the latter).  At high pressures the continuum hydrogen opacity and/or the cloud opacity tends to fill in the opacity trough at the minimum of the water opacity in $H$ band.  Since the photosphere of lower gravity objects at fixed effective temperature is at lower pressures, the $\rm H_2$ and cloud opacity 
is somewhat less important allowing the angular shape of the water opacity to more strongly control the emergent flux (see Figure \ref{models} and Figure 6 of \citet{Ric11}).  

Thus we find that the shape of the $H$ band spectrum is responsible for pulling the preferred model fits to low gravity and low effective temperature. Weaker methane bands at lower $\log g$ in this $T_{\rm eff}$ range also push the fit to lower gravity.  The greater number of datapoints in the spectra overwhelms the photometric data which is why the contours for the best overall fit lie outside of the accepted luminosity range.  As we discuss in Section 5.1 our preferred interpretation is that none of our current models match the true composition, mass, and age of this planet.

The model which best fits the photometry alone in the top panel of Figure \ref{planet_b} fits the $YJHK$ and [3.3]-$\rm \mu m$ (but not the revised \citet{Ske12} [3.3]) photometry to within $1\sigma$.  The model is too bright at $L^\prime$ and $M^\prime$.  The photometry plus spectrum fit features a methane band head at $2.2\,\rm \mu m$ that is too prominent, even with $\log K_{zz} = 4$.    
Both sets of solutions,  are inconsistent with the accepted age of the the star.  The lower mass solution would imply very young ages for the planet, well below 30 Myr. Conversely the higher mass range implies ages in excess of about 300 Myr.   Thus along with the discarded low mass fit  the photometry-only, higher mass fit is problematical since the mass conflicts with the constraints discussed in Section 4.1
 
\subsubsection{HR 8799c}

For planet c there is no available spectroscopy and we fit only to the photometry.  The formal best fitting solution yields $T_{\rm eff}=980\pm70\,\rm K$ and $\log g = 4.33 \pm 0.28$ for a mass of $15\pm 8\,\rm M_{\rm J}$.  However in both the contour diagram (Figure \ref{contour}) and the histogram (Figure \ref{hist}) we find two islands or clusters of acceptable fits, one at higher gravity and effective temperature, and one with lower values for both.  The high mass solution lies 
at masses greater than $20\,\rm M_J$ and $T_{\rm eff}\sim 1100\,\rm K$.  Such models are consistent only with ages around 300 My, well in excess of the preferred age  range for the primary and the dynamical constraints on the mass.  The second island of acceptable fits lies at $\log g \sim 4.25$ and $T_{\rm eff}\sim 950\,\rm K$.  Figure \ref{planet_c} illustrates the spectra for the best fitting model from each case. 
The lower mass  model has $\log g = 4.25$,  $f_{\rm sed}=2$, and $\log K_{\rm zz}=4$, implying $M\approx10\,\rm M_J$ which is consistent with the dynamical  mass constraint and represents our preferred solution and is listed in Table 1.  The age predicted by the evolution of these models is about  160 Myr, consistent with the asteroseismological age constraint but not the generally favored range of 30 to 60 Myr.  However models with modestly lower gravity and slightly smaller masses also fall within the $1\sigma$ contours seen in Figure \ref{contour}  do lie within this age range.  

The cooler model fits most of the photometric points to within 2$\sigma$ or better, but varies most significantly from the data at $[3.3]\, \rm \mu m$ and $L^\prime$, which perhaps imply that despite the disequilibrium chemistry the models have too much methane.  The lower gravity solutions differ from the high gravity ones most prominently in the red side of $K$ band (where the cooler model has a much more prominent methane band head) and at 3 to $4\,\rm \mu m$.  By constraining the methane band depth in the $K$ band and, to a lesser extent,  in the $H$ band, spectroscopy has the potential to distinguish between these two cases.  The shape of $H$ band (Figure \ref{models}) can also serve as a gravity discriminator with a more triangular shape indicating lower gravity.

\subsubsection{HR 8799d}

Because of larger observational error bars, the model fits for the innermost of the three planets considered here are the most uncertain.  As seen in Figure \ref{contour} the best fitting models allow masses ranging from 5 to $60\,\rm M_J$ and  $T_{\rm eff}$ between 900 and 1200 K.  However the very best fitting models favor solutions  with $\log g$ around 4.25 to 4.50 and $T_{\rm eff} = 1000\,\rm K$ yielding a mass of 10 to $20\,\rm M_{\rm J}$.   As with planet c such a solution is consistent with the dynamical constraint but not the age constraint.  Also as with planet c the lower end of this mass range offers marginally poorer fits that nevertheless still lie within the $1\sigma$ contour and that do satisfy the age constraint.  The best fitting spectrum is shown in Figure  \ref{planet_d}.

\section{Discussion}
\subsection{Implied Masses and Ages}
To summarize our findings from the previous section, each of the three planets considered presents a different challenge to characterize.  Some model fits to planets c and d imply implausibly large masses or ages but other acceptable fits satisfy all of the available constraints.  Both c and d can be characterized as having masses as low as 7 to $8\,\rm M_{\rm J}$, $T_{\rm eff}=1000\,\rm K$, and $f_{\rm sed}=2$ which implies ages of around 60 Myr, within the most commonly cited age range of the primary.  Some better fitting models have slightly larger masses ($10\,\rm M_{\rm J}$) and ages (160 Myr). This age is greater than the range of ages typically quoted for the primary star of 30 to 60 Myr, although it is within the range permitted by the asteroseismology.  Evolution models starting from a cooler initial state than the hot-start models would reach these effective temperatures and gravity at a younger age than 160 Myr and be more in accord with the usual age range.  

For planet b none of the models are satisfactory.  Since we do not allow arbitrary radii models to fit the data (with the exception of the lowermost panel in Figure \ref{planet_b}), we cannot invoke what we judge to be unphysical radii to produce acceptable fits.  The solution which best fits the photometry alone has $M=26\,\rm M_{\rm J}$, $T_{\rm eff}=1000\,\rm K$, and $f_{\rm sed}=2$, but this mass clearly violates the constraints discussed in Section 4.1.  A fit to the entire spectral and photometric dataset results in  $M\approx 3\,\rm M_{\rm J}$, $T_{\rm eff}=750\,\rm K$, and $f_{\rm sed}=1$.  However we discard this model as discussed in Section 4.4.1.   This effective temperature is cooler than favored by \citet{Bar11a} and \citet{Cur11} but is comparable to that found by \citet{Mad11}.  

The most likely explanation for the difficulty in fitting this object is that one of the assumptions of the modeling is incorrect.  \citet{Bar11a} speculate that a super-solar atmospheric abundance of heavy elements might explain the departures of the data from the models.  Indeed all of the atmospheres of solar system giant planets are enhanced over solar abundance with a trend that the enhancement is greater at lower masses.  For example Saturn's atmosphere is enhanced in methane by about a factor of ten while Jupiter is only a factor of about three (see \citet{Mar07} for a review).  The available data on exoplanet masses and radii suggest that lower mass planets are more heavily enriched in heavy elements than higher mass planets \citep{Mil11}. If the mass of HR 8799b is intermediate between our two sets of best fits, for example with a mass near 6 or $7\,\rm M_J$, as favored by the discovery paper, and if atmospheric abundance trends are similar in the HR 8799 system to our own, then it may not be surprising if planet b has different atmospheric heavy element abundances than c and d.  We will consider non-solar abundance atmosphere models in a future paper.    The full range of model phase space has certainly not yet been explored.

Overall we find that a consistent solution can be found for planets c and d in which both have similar masses and ages.  This is essentially the solution favored by the discovery paper \citep{Mar08} and is within the ranges of favored solutions presented by \citet{Cur11} and \citet{Mad11}.   However we differ from some of  these previous studies in our finding  that the radii for planets b and c that are fully consistent with that expected for their individual masses.  Unusual radii are not required. 

\subsection{Bolometric Luminosities}

The distance to HR 8799 has been measured as $d=39\pm1.0\,\rm pc$
\citep{van07} and thus the bolometric luminosity of each planet can
be computed from the observed photometry.  In the discovery paper,
\citet{Mar08} compare the photometry available at that time to models
and brown dwarf spectra and report the now commonly cited results $\log
L_{\rm bol} / L_\odot = -5.1\pm0.1$ for planet b and $-4.7\pm0.1$ for
c and d.   

Since the work of \citet{Mar08}, the photometry of the three planets
has been expanded to cover the SED from $\sim 1$--4.8$\,\mu$m.  This
better constrains $L_{\rm bol}$  as $\sim$80\% of the flux is emitted at these wavelengths.
In principle, the bolometric luminosity can be obtained by fitting
synthetic photometry to the observations, with a scaling factor
chosen to minimize  the residuals. The integrated scaled flux of
the model and the known distance gives $L_{\rm bol}$ \citep{Mar08}.
The fitted model thus provides an effective bolometric correction to
the photometry by approximating the flux between the
photometric bands.  The scaling factor corresponds to
$(R/d)^2$, where $R$ is the radius of the planet. The optimized
scaling thus
corresponds to an optimization of the radius independent of the
physical radius of the planet. As is well known, this results in radii
for the HR 8799 planets that are considerably smaller
than be accounted for with the evolution models (Section 1.1).  The
approach can also lead to unrealistic bolometric corrections if the
fitted $\teff$ deviates too far from the actual value.

To circumvent this difficulty, here we determine $L_{\rm bol}$ by
using the radius obtained from our evolution sequences, which is
consistent with our approach to fit the photometry. Of course such
theoretical radii have their own uncertainty, including a dependence
at young ages -- particularly below 100 Myr -- on the initial conditions
\citep{Bar02,Mar07a,Spi12}.   We neglect the dependence on initial
conditions since planets forming in  the `cold-start' calculation of
\citet{Mar07a} never  get as warm or as bright as the HR 8799
planets.  Intermediate cases, such as explored by \citet{Spi12} are
possible, but we set those aside for now.  Our approach, however, does
eliminate unphysical solutions by constraining the radius to
reasonable values (in excess of $1\,\rm R_J$). Thus, for each fitted
model $(\teff, \log g)$ we obtain a $L_{\rm bol}$ from the radius $R(\teff, \log g)$ 
obtained with the evolution\footnote{With ``perfect'' atmosphere and evolution models the two
methods would give identical results.}.
The resulting probability distributions of $L_{\rm bol}$ for each
planet are shown in Fig. \ref{hist}, along with the mean value and
dispersion of each distribution.

Our fits are based on a model grid with spacing of 50$\,$K and 0.25$\,
$dex in $\teff$ and $\log g$, respectively, which introduces an
additional uncertainty intrinsic to the fitting procedure of
about half a grid spacing, or $\pm25\,$K and $\pm0.13\,$dex.  We
derive the corresponding uncertainty in $L_{\rm bol}$ as follows.  The
bolometric luminosity is given by
$$ L_{\rm bol}=4\pi R^2 \sigma T_{\rm eff}^4 = {4\pi GM_\odot \sigma
\teff^4\over g}\Bigl({M \over M_\odot}\Bigr),$$
where the symbols have their usual meaning.  From the cloudy evolution
of \citet{Sau08}, we find an approximate relation for $M(\teff,\log g)
$ in the range of $\teff$ and mass of interest:
 $$  \log {M \over M_\odot} = 0.746\log g + {\teff \over 5090} - 5.35,$$
 where $\teff$ is in $K$ and $g$ in cm$\,$s$^{-2}$. Thus,
$$  \log L_{\rm bol} = 4\log\teff  + {\teff \over 5090} - 0.254\log g +
A,$$
where $A$ is a constant. With the grid spacing uncertainties given
above, we find $\Delta \log L_{\rm bol}= \pm0.054$, which we round up
to 0.06.
Combining quadratically this uncertainty with the dispersion in
$L_{\rm bol}$ found in our fits (Fig. \ref{hist}), we find the luminosity
for planet b to be $\log L_{\rm bol}/L_\odot=-4.95\pm0.06$, $-4.90 \pm
0.10$ for planet c\footnote{Note that the dispersion for planet c is
non-Gaussian (Fig. \ref{hist}).}, and $-4.80 \pm 0.09$ for planet d. These
values are consistent with those reported by \citet{Mar08}
although they are 0.1$\,$dex brighter for planet b, 0.2$\,$dex fainter
for planet c, and 0.1$\,$dex fainter for planet d. The quoted
uncertainties are lower limits of course, since they do not account for
obvious systematic errors in the models (Figures \ref{planet_b}--\ref{planet_d}).

\subsection{Cloud Properties}

Although there is a dispersion in the best fitting $\log g$ and $T_{\rm eff}$, essentially all of the acceptable fits require a cloud sedimentation
efficiency of $f_{\rm sed}=2$.  As shown in Figure 1 this value is typical of the best fitting parameters for most field L dwarfs we have previously studied \citep{Cus08, Ste09}.  The persistence of clouds to lower effective temperatures at low gravity is also apparent from this figure.  By 1000 K most field dwarfs with $\log g \ge 5$ have already progressed to $f_{\rm sed} \ge 4$ whereas clouds persist much more commonly among lower gravity objects down to 1000 K.  By very cool effective temperatures, however, the silicate and iron clouds have certainly departed from view as demonstrated by the one $\log g = 4$, $T_{\rm eff}\sim 500\,\rm K$ object (ULAS J133553.45+113005.2, \citep{Bur08, Leg09}). 

As Figure 1 attests, the cloud in planets b, c, and d are unusual not so much for their global characteristics (the same cloud model that describes L dwarf clouds fits them as well), but rather for their persistence.  At $f_{\rm sed} = 2$ there are three  field objects with $T_{\rm eff} \le 1200\,\rm K$.  These objects are 2MASS 0825196+211552 \citep{Kir00}, SDSS 085758+570851 \citep{Geb02}, and  SDSS J151643.01+305344.4 (\citet{Chi06}; hereafter SDSS 1516+30).  Their infrared spectral types are L6, L8, and T0.5 and the first two are both redder in $J-K$ than is typical for those spectral types \citep{Ste09}.   

Figure 3  compares some of the model silicate cloud properties of  a low gravity planet    with  models for   field L6 and T0.5 objects.  As expected from the discussion in Section  2.2, the lower gravity model is marked by a larger particle size than the higher gravity models,  and the column optical depth of the silicate cloud in all three objects ends up being very similar.  More importantly the range of cloud optical depths that lie in the near-infrared photosphere are similar for all three objects.  Thus a low gravity ($\log g = 3.5$) object with $T_{\rm eff} = 1000\,\rm K$ ends up with cloud opacity that is very similar to a high gravity ($\log g = 5.5$) object with $T_{\rm eff} = 1200\,\rm K$ and consequently similar spectra and colors. Indeed \citet{Bar11a} has already noted the similarity of SDSS 1516+30 to HR 8799b.
  This congruence between lower gravity and higher gravity models led to the initial surprise that the apparently cool planets seem to have clouds reminiscent of higher gravity--and warmer--L dwarfs.

The relative contribution of clouds to the opacity in individual photometric bands is depicted in Figure \ref{contrib}.  This figure presents contribution functions for the $J$, $H$, $K$, $L^\prime$, and $M^\prime$ bands for six different combinations of gravity, effective temperature, and cloud treatment.  The contribution functions illustrate the fractional contribution to the emergent flux as a function of pressure in the atmosphere.   In a cloud-free, $T_{\rm eff}=1000\,\rm K$, $\log g = 5.0$ atmosphere (left panel, Figure \ref{contrib}a) the $L^\prime$ flux emerges predominantly near $P=0.6\,\rm bar$ while the $J$-band flux emerges from near 8 bar.  The contribution functions do not account for the effect of cloud opacity, but rather show for each case where the flux would emerge from for that particular model if there were no clouds.

The center two panels of Figure \ref{contrib}a and b illustrate the vertical location of the cloud layers for both $f_{\rm sed}=1$ and 2.  The $f_{\rm sed}=2$ clouds are thinner and the cloud base is deeper since these less cloudy atmospheres are cooler than the $f_{\rm sed}=1$ case, as seen in the right hand panels.  If the cloud deck lies above or overlaps the plotted contribution function of a given band then the emergent flux in that band will be strongly affected by the presence of the cloud.  The figure makes clear that regardless of gravity thicker clouds impact more of the emergent spectra than thinner clouds.  Clouds described by $f_{\rm sed}=2$ strongly impact $J$, $H$, and $K$ bands, but are less important at $L^\prime$, and $M^\prime$.  We conclude that at least for the effective temperature range inhabited by HR 8799 b, c, and d that clouds are most strongly impacting the observed spectra at wavelengths shorter than about $2.5\,\rm \mu m$ while the longer wavelength flux is primarily emerging from above the cloud tops.  Figures such as this illustrate the value multi-band photometry has in both constraining not only the total emergent flux, but also the vertical structure of the clouds.

\subsection{Evolution with a gravity-dependent L to T transition}

The growing evidence that the cloudy to cloudless transition in
field brown dwarfs depends on gravity (\S2.1) is complemented by
the published analyzes of the HR 8799 planets (including the present work)
which all indicate that their atmospheres are cloudy and that they have $\teff$ 
well below the estimated $\sim 1400 \,$K limit of the L dwarf sequence. Thus, it appears that the
atmospheres of lower gravity dwarfs and of imaged exoplanets retain their clouds to lower $\teff$,
which is supported by simple cloud model arguments (\S2.2). As we have argued,
this is the simplest interpretation of the fact that the HR 8799 planets have
$\teff$ typical of cloudless T dwarfs but have evidently cloudy atmospheres.
How is the evolution of brown dwarfs across the transition from cloudy to clear atmosphere affected?

The atmosphere of a brown dwarf largely controls its evolution because it acts as a
surface boundary condition for the interior.  A more opaque atmosphere (more
clouds, or higher metallicity, for instance) slows the escape of radiation and increases
the cooling time of the interior. 
\citet{Sau08} looked at the evolution of brown dwarfs across the transition
by assuming that the atmosphere was cloudy ($\fsed=2$) down to $\teff=1400\,$K,
and clear below 1200$\,$K, with an linear interpolation of the atmospheric boundary condition
in the transition regime.  This effectively corresponds to increasing the sedimentation efficiency across
the transition, one of the proposed explanations for the cloud clearing (\S2.1). By converting 
the evolution sequences to magnitudes using
synthetic spectra ($f_{\rm sed}=1$ for cloudy atmospheres, and $f_{\rm sed}=4$ for
``clear'' atmospheres\footnote{These are not fully consistent with the values used
for the evolution, but the effect on the evolution of this small difference in $\fsed$ is small.}) 
a good match to the near-infrared color
magnitude diagrams of field dwarfs was found from the cloudless late M dwarfs, along the cloudy 
L dwarf sequence, across the L/T transition and down to late T dwarfs.

We now extend this toy model to include a gravity-dependent range of $\teff$ for the transition
to explore the consequences, at the semi-quantitative level, on the cooling tracks of
brown dwarfs and exoplanets.  In view of the success obtained for field dwarfs (of relatively
high gravity) with the \citet{Sau08} toy model, and the requirement that the lower gravity HR 8799 planets
be cloudy at $\teff \sim 1000\,$K, we define the transition region to be $\teff = 1400$ to 1200$\,$K
at $\log g=5.3$ (cgs) and 900 to 800$\,$K at $\log g=4$ with a linear interpolation in between
(Fig. \ref{trans_gdep}). The cloudy boundary condition above the transition is based on our $\fsed=2$
atmosphere models, and our cloudless models below the transition, as in \citet{Sau08}. Synthetic
magnitudes are generated from the cooling tracks using our new $\fsed=1$ and cloudless atmosphere
models \citep{Sau12}.

The resulting cooling tracks of two low-mass objects of 5 and 20$\,$M$_{\rm J}$ are shown in 
Fig. \ref{evol_gdep} where the same calculation, but based on a fixed $\teff$ transition 
(Fig. \ref{trans_gdep}) is also displayed for comparison. It is immediately apparent that these low-mass 
objects, which retain their clouds to lower $\teff$ ($\sim 850\,$K for 5$\,$M$_{\rm J}$ and $\sim 1050\,$K
for 20$\,$M$_{\rm J}$) with the prescribed gravity-dependent transition evolve along the L dwarf
sequence longer and reach the region of the color-magnitude diagram occupied by the HR 8799 planets before they turn to
blue $J-K$ colors as the cloud clears. Also remarkable
is that in the transition region where the $J-K$ color changes from $\sim 2$ to $\sim 0$, the
low mass object is {\it fainter} in $K$ than the higher mass object, the reverse of the situation
for a transition that is independent of $\teff$. This effect persists up to a cross over mass of
$\sim 60\,$M$_{\rm J}$ above which the trend reverses (Fig. \ref{trans_gdep}).  This implies that
low mass objects that are in the transition region should appear below (i.e. be dimmer) the field
T0--T4 dwarfs, perhaps by up to 1--2 magnitudes.  We note that the pile up of objects in the transition
region reported in \citet{Sau08} still occurs in this new calculation but it is more spread out in $\teff$,
as would be expected from the broader span of the transition in $\teff$ (Fig. \ref{trans_gdep}).

We emphasize that this evolution calculation is a toy model that has been loosely adjusted to account for
limited observational constraints. It reveals trends but is not quantitatively reliable.  In particular,
we have had to use $\fsed=1$ to match the near infrared colors of the HR 8799 planets while
our best fits give $\fsed=2$ for all three planets. This reflects the fact that the models
give different best-fit parameters when applied to a subset of the data, a well-known difficulty with 
current models \citep{Cus08,patience12}.

\subsection{Mixing}

Given the discussion in Section 1.3 regarding the prevalence of atmospheric mixing resulting in departures from chemical equilibrium in solar system giants and brown dwarfs, it is not surprising that mixing is also important in warm exoplanet atmospheres as well.  \citet{Bar11a} discuss the influence of non-equilibrium chemistry at low gravity and find that the $\rm CO/CH_4$ ratio can become much larger than 1 in the regimes inhabited by the HR 8799 planets.  Also \citet{Bar11b} found non-equlibrium chemistry was likely important in 2M1207b.

We find that all of the best fitting models for each planet, b, c, and d, include non-equilibrium chemistry.  Within our limited grid with $K_{zz}= 0$ and
$10^4\,\rm cm^2\,s^{-1}$, the latter choice was strongly preferred in all cases providing yet another indication of the importance of chemical mixing in substellar atmospheres.  This also suggests that a fuller range of models with a greater variety of eddy mixing strengths should be considered in future studies to better constrain this parameter.

\subsection{Mechanism for Gravity Dependent Transition}

In Section 2.2 we demonstrated that the effect of a given cloud layer, all else being equal, is greater in a lower mass extrasolar giant planet than in
a more massive brown dwarf of the same effective temperature.  If we add effective temperature as a variable then we find that a cooler low mass object can have clouds comparable to a warmer high mass object.  Such a congruence is empirically demonstrated by the similar spectra of SDSS 1516+30  and HR 8799 b (as originally noted by \citep{Bar11a}).  The former is a $\sim70\,\rm M_J$, 1200 K field L dwarf while the latter is plausibly a few Jupiter mass, 1000 K young gas giant planet (although the modeling discussed here does not select this solution).  Likewise in Section 5.4 our simple evolution calculation with a gravity-dependent L to T type transition temperature illustrates
that the location of young objects on the color magnitude diagram can be understood if clouds remain to lower effective temperatures at lower gravity.  The fact that such behavior is dependent upon gravity is not in itself surprising as a lower gravity would be expected to alter its behavior.   However the specific question remains, what is the specific mechanism that results in lower mass objects making the L to T type spectral transition at lower effective temperatures than higher mass objects?  In this section we offer some speculation while recognizing that a serious analysis is beyond the scope of this paper.

A possible contributing factor might be found in the relative positions of  the convection zone and the photosphere as a function of gravity (a point also raised in \citet{Bar11a, Bar11b} and  \citet{Ric11}.  To illustrate this effect in Figure \ref{contrib} the 
 contribution functions for different bandpasses are shown for two different gravities.  At $T_{\rm eff}=1000\,\rm K$ for moderately cloudy ($f_{\rm sed}=2$) atmospheres the convection zone, regardless of gravity, penetrates into the   cloud layers that control the $J$ and $H$ band fluxes.  For cloudless atmospheres, however, the convection zone for the high gravity case is quite deep ($P>20\,\rm bar$), well below even the region probed by the $J$ band (Figure \ref{contrib}a).  At lower gravity, however, the convection zone penetrates higher into the atmosphere to much lower pressure, overlapping the $J$ band contribution function (Figure \ref{contrib}b).   If we imagine that a given patch of atmosphere begins to clear, perhaps because of more efficient local sedimentation, in the high gravity case the removal of cloud opacity leads the atmosphere  to become radiative and more quiescent, favoring 
particle sedimentation relative to convective mixing and enlarging what began as a localized clearing.  At low gravity however the removal of cloud opacity does not as dramatically push the atmosphere to a quiescent state.  Thus convection continues to loft cloud particles and the local clearing fills back in.  Only when the clear atmosphere convection zone lies very deep do the clouds dissipate.  Since low gravity atmospheres are more opaque than high gravity ones this process of the growth of clearings begins at lower effective temperature at lower gravity.

Another possibility is that detached convection zones play a role in hastening the L to T transition.  Within some effective temperature ranges there are two atmospheric convection zones, one deeply seated and a detached zone that is separated by a small radiative zone.  This can be seen in the $f_{\rm sed}=1$ temperature profiles in Figure \ref{contrib}.  \citet{Bur06} and \citet{Wit11} have speculated that the interplay of dynamical and cloud microphysics effects that may occur when the intermediate radiative zone forms or departs may play a role in the transition.  Perhaps at some effective temperature threshold particles forming in the upper convective zone grow large enough that they fall all the way through the cloud base and the intermediate radiative zone before they completely evaporate. Depending on the efficiency of mixing in the radiative zone this could result in a net transport and sequestration of condensate away from the near-infrared photosphere.  \citet{Wit11} discuss a similar idea of the convection ``fanning'' the fall of particles away from the upper zone.  As seen in Figure \ref{contrib}, however, for both the $f_{\rm sed}=2$ and the cloudless case there is only one convection zone, so the potential for multilayered convection is less compelling in this case.  Nevertheless such mechanisms require more sophisticated modeling to ascertain how they might be affected by gravity and effective temperature.

Arguments such as these that are based upon 1D radiative convective models only scratch the surface of the underlying complex dynamical problem.  For example \citet{Fre10} performed two-dimensional radiation hydrodynamic simulations of brown dwarf atmospheres to study the effects of clouds on atmospheric convection.  They found that atmospheric mixing driven by cloud opacity launches gravity waves that in turn play a role in maintaining the cloud structure.  The Freytag et al.\ study considered a domain a few hundred kilometers wide by about 100 km deep and only investigated a single gravity ($\log g=5$) so how such effects might vary with gravity is not yet known.  Furthermore how the local clouds might interact with the very large scale planetary circulation has not been explored.  Perhaps clouds form holes or otherwise dissipate only when most of the cloud optical depth lies deeply enough to be strongly influenced by global atmospheric circulation.  Large scale global dynamical simulations that capture the relevant physics of particle and energy vertical and horizontal transport are likely required to fully describe the L to T transition mechanism.

\subsection{Future}
 
Our experience in fitting the spectra of planet b in particular points to the importance of spectra in the analysis. Adding the $H$ and $K$ band spectra to the analysis results in much lower preferred masses than fitting photometric data alone.  Thus we expect that additional spectral data will further inform future model fits.  

As noted in Section 2.1 one hypothesis for the nature of the L to T transition is that it involves partial clearing of the assumed global cloud cover.  It is possible that models which include partial cloudiness may better describe the observed flux and \citet{Cur11} have explored this possibility.  Given
the limited data available today we feel the addition of another free model parameter is premature and in any event we have found that brown dwarfs with partial cloud cover have an overall near-infrared spectrum that resembles a homogeneous dwarf with a thinner, homogenous global cloud \citep{Marl10}.

Another method for characterizing these planets and probing atmospheric condensate opacity in self-luminous planets is by polarization \citep{Marl11, dek11}.  \citet{Marl11} found that rapidly rotating, homogenously cloud-covered planets may be sufficiently distorted to show polarization fractions of a few percent if they are relatively low mass.  \citet{dek11} found that even when partial cloudiness is considered much larger polarization fractions are unlikely.  However if this level of polarization could be measured in one of the HR 8799 planets this would confirm the presence of clouds and also place an upper limit on the planetary mass.   Objects in this effective temperature range (near 1000 K) and with $\log g > 4$ are predicted to exhibit polarization well below 0.2\%.  Both SPHERE and GPI have polarization imaging modes, but it is not clear if they would have sufficient sensitivity to place useful upper limits on the HR 8799 system.

\section{Conclusions}

We have explored the physical properties of three of the planets orbiting HR 8799 by fitting our standard model spectra to the available photometry and spectroscopy.  Unlike some previous studies we have required that models with a given $\log g$ and $T_{\rm eff}$ have a corresponding radius that is calculated from a consistent set of evolution models.  While the radii of the planets are not variables, we do include two other free parameters: the cloud sedimentation efficiency $f_{\rm sed}$ and the minimum value of the atmospheric eddy mixing coefficient $K_{zz}$.

In agreement with all previous studies we find that the atmospheres of all three planets are cloudy, which runs counter to the expectation of conventional wisdom given their relative low effective temperature.  However as we argue in Sections 2.1 and 2.2, finding clouds to be present at lower effective temperatures in lower gravity objects is fully consistent with trends already recognized among field L and T dwarfs and from basic atmospheric theory.  We uniformly find that the best fitting value of the sedimentation efficiency $f_{\rm sed}$ is, in essentially all cases, 2, which is typical of the value seen in pre-L/T transition field L dwarfs (Fig. 1) \citep{Cus08, Ste09}.  In agreement with \citet{Bar11a} we thus find that the clouds in these objects are neither ``radically enhanced'' \citep{Bow10} nor representative of a ``new class'' \citep{Mad11} of atmospheres. 

As have some previous authors \citep{Bar11a, Bar11b} we find that eddy mixing in nominally stable atmospheric layers is an important process for altering the chemical composition of all three planets.  While we have not carried out a comprehensive survey of non-equilibrium models, we find that values of the eddy mixing coefficient near $\log K_{zz}\sim4$ generally fit the available data better than models that neglect mixing.  Such values are typical of those found for field L and T dwarfs \citep[e.g.,][]{Ste09} and the stratospheres of solar system giant planets (e.g., see the detailed discussion for Neptune in \citet{Bis95}).

The best fitting values for the primary model parameters $\log g$ and $T_{\rm eff}$ are less secure.  For HR 8799 b the inclusion of the $H$ and $K$ band spectra of \citet{Bar11a} drive our fits to low masses of $\sim 3\,\rm M_J$ and effective temperatures, a solution which we discard as discussed in Section 4.1.1.  The photometry alone favors much higher masses, $\sim 25\,\rm M_J$ that are apparently ruled out by dynamical considerations.  Thus we find no plausible model that fits all of the accepted constraints.  Fits for the planets c and d likewise  generally favor higher masses, although there are some solutions that are consistent with masses near or below $\sim 10\,\rm M_J$ with ages consistent with the available constraints. For all three planets the photometry predicted by the best fitting model is generally consistent with the observed data within 1 to 2 standard deviations.    We stress that all of these fits have radii that are appropriate for the stated effective temperature and gravity.

In conclusion the modeling approach that has successfully reproduced the spectra of field L and T dwarfs seems to also be fully applicable to the directly imaged planets.  Nevertheless a larger range of model parameters, including non-solar metallicity, must be explored in order to fully characterize these objects as well as the planets yet to be discovered by the upcoming GPI, SPHERE, and other coronagraphs.
 
\section{Acknowledgements}

We thank Travis Barman and Bruce Macintosh for helpful conversations and Travis Barman for a particularly helpful review.  
This material is based upon work supported by the National Aeronautics and Space Administration  through the Planetary Atmospheres and Astrophysics Theory Programs as well as the Spitzer Space telescope Theoretical Research Program.  This research was supported in part by an appointment to the NASA Postdoctoral Program at the Jet Propulsion Laboratory, administered by Oak Ridge Associated Universities through a contract with NASA.  Based in part on data collected at Subaru Telescope, which is operated by the National Astronomical Observatory of Japan.
Observations used here were obtained at the MMT Observatory, a joint facility of the University of Arizona and the Smithsonian Institution.
Some of the data presented herein were obtained at the W.M. Keck Observatory, which is operated as a scientific partnership among the California Institute of Technology, the University of California and the National Aeronautics and Space Administration. The Observatory was made possible by the generous financial support of the W.M. Keck Foundation.

\vfill\eject

\vfill\eject
\begin{deluxetable}{clcccccc} 
\tablecolumns{8} 
\tablewidth{0pc} 
\tablecaption{Summary of Derived Planet Properties} 
\tablehead{ 
\colhead{Planet} & \colhead{Ref.\tablenotemark{1}}   & \colhead{$M$ ($M_{\rm Jup}$)}   & \colhead{$\log g$} & $T_{\rm eff}\,\rm (K)$ & 
$R (R_{\rm J}$) & age (Myr)&$\log L_{\rm bol}/L_\odot$  }
\startdata 
b\tablenotemark{2} & B11a & $0.1 - 3.3$ &$3.5\pm0.5$ & $1100\pm 100$ &0.63 - 0.92 & $30 - 300$&$-5.1\pm0.1\tablenotemark{3}$ \\
   & C11& $5 - 15$ &$4-4.5$ & $800-1000$ &\nodata & $30 - 300$ \\
   & G11  & 1.8 &$4$ & $1100$ & 0.69 & \nodata \\
   & M11  & $2-12$ &$3.5-4.3$ & $750-850$ &\nodata & $10-150$ \\
    \cline{2-8}
   & M12\tablenotemark{4} &  26 & 4.75  & 1000    & 1.11& 360 &$-4.95\pm0.06$ \\ 

 \cline{1-8}
c & C11  & $7 - 17.5$ &$4-4.5$ & $1000-1200$ & \nodata & $30 - 300$&$-4.7\pm0.1\tablenotemark{3}$ \\
   & G11  & 1.1 &$3.5$ & $1200$ &0.97 &\nodata  \\
   & M11 & $7-13$ &$4-4.3$ & $950-1025$ & \nodata & $30-100$ \\
    \cline{2-8}
   & M12 & 8 -- 11 &$4.1\pm0.1$ & $950\pm 60$  &1.32 -- 1.39 &40 -- 100&$-4.90\pm0.10$  \\

 \cline{1-8}
 d & C11  & $5 - 17.5$ &$3.75-4.5$ & $1000-1200$ &\nodata & $30 - 300$&$-4.7\pm0.1\tablenotemark{3}$ \\
   & G11  & 6 &$4.0$ & 1100 &1.25 & \nodata \\
   & M11  & $3-11$ &$3.5-4.2$ & $850-1000$ &\nodata & $10-70$ \\
    \cline{2-8}
   & M12 &  $8-11$ & $4.1\pm0.1$& $1000\pm 75$  & 1.33 -- 1.41&  30 -- 100&$-4.80\pm0.09$\\

\enddata 

\tablenotetext{1}{B11a=\citet{Bar11a}; C11=\citet{Cur11}; G11 = \citet{Gal11}; M11=\citet{Mad11}; M12=this work.}
\tablenotetext{2}{Parameters derived by \citet{Bow10} are not listed because of very large scatter depending upon various assumptions.}
\tablenotetext{3}{\ Luminosity from \citet{Mar08}.}
\tablenotetext{4}{For b this is the formal best fit single model to the photometry alone, for c and d these are the preferred solution ranges as discussed in the text.  The b fit is incompatible with the generally accepted constraints as discussed in the text.  Formal solutions are shown in Figure \ref{hist}.}
\end{deluxetable}

\begin{deluxetable}{cccl} 
\tablecolumns{4} 
\tablewidth{0pc} 
\tablecaption{Photometric Data for the HR 8799 Planets} 
\tablehead{ 
\colhead{Planet} & \colhead{Band}   & \colhead{Abs. Mag.}   & \colhead{Ref.\tablenotemark{1}}}
\startdata 
b & Subaru-$z$  & $18.24 \pm 0.29$ & C11 \\
   & $J$  & $16.52 \pm 0.14$ & C11  \\
   & $H$  & $14.87 \pm 0.17$ & M08  \\
   & $K_s$ & $14.05\pm0.08$ & M08  \\ 
   & [3.3]  & $13.96\pm0.28$ & C11  \\
   & $L^\prime$  & $12.68\pm0.12$ & C11  \\
   & $M^\prime$ & $13.07\pm0.30$ & G11  \\
 \cline{1-4}
c & Subaru-$z$  & $>16.48$ & C11 \\
   & $J$  & $14.65 \pm 0.17$ & M08  \\
   & $H$  & $13.93 \pm 0.17$ & M08  \\
   & $K_s$ & $13.13\pm0.08$ & M08  \\ 
   & [3.3]  & $12.64\pm0.20$ & C11  \\
   & $L^\prime$  & $11.83\pm0.07$ & C11  \\
   & $M^\prime$ & $12.05\pm0.14$ & G11  \\
 \cline{1-4}
d & Subaru-$z$  & $>15.03$ & C11 \\
   & $J$  & $15.26 \pm 0.43$ & M08  \\
   & $H$  & $13.86 \pm 0.22$ & M08  \\
   & $K_s$ & $13.11\pm0.12$ & M08  \\ 
   & [3.3]  & $>11.63$ & C11  \\
   & $M^\prime$ & $11.67\pm0.35$ & G11  \\
\enddata 
\tablenotetext{1}{C11=\citet{Cur11}\\ M08=\citet{Mar08} \\ G11=\citet{Cur11}}
\end{deluxetable} 

\clearpage
\begin{figure}
\includegraphics[angle=0.0,scale=.15]{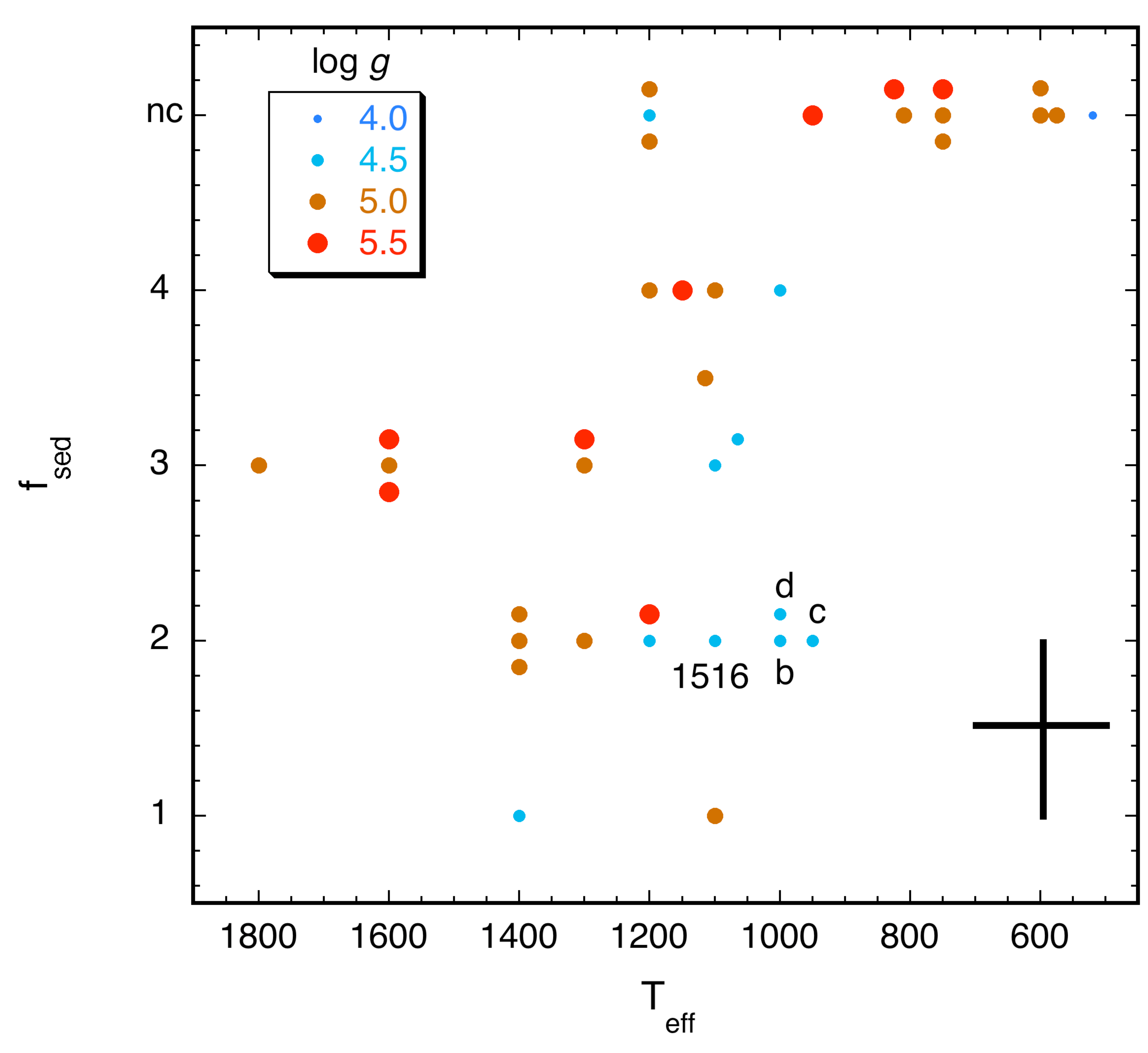}
\caption{Model parameters $f_{\rm sed}$ and $T_{\rm eff}$ as derived by various applications of Marley \& Saumon atmosphere and evolution models.  Size of dot reflects derived $\log g (\rm cm\,s^{-2})$ and `nc' denotes cloudless models (note that `nc', which corresponds to $f_{\rm sed}\rightarrow \infty$, is arbitrarily plotted at $f_{\rm sed}=5$).  Points which would otherwise overlap are slightly offset vertically and the $T_{\rm eff}$ values decrease to the right to suggest evolution in time.  The points for HR 8799 c and d from the analysis here are labeled with planet designator.   Remaining  points are from \citet{Geb01, Mai07, Leg07a, Leg08, Geb09, Leg09, Ste09, Mai11} although fits to unresolved binaries and objects with very poorly constrained properties (e.g., Gl 229 B with $\log g$ uncertain by a full dex) are excluded. SDSS 1516+30 is denoted by `1516'.  The cross denotes size of the typical uncertainties in the model fits which are usually $\pm 100\,\rm K$ in effective temperature, $\pm0.25\,\rm dex$ in $\log g$, and $\pm 0.5$ in $f_{\rm sed}$, although the uncertainty analysis is not uniform across the various sources. }
\label{fig1}
\end{figure}

\clearpage
\begin{figure}
\includegraphics[angle=0.0,scale=0.7]{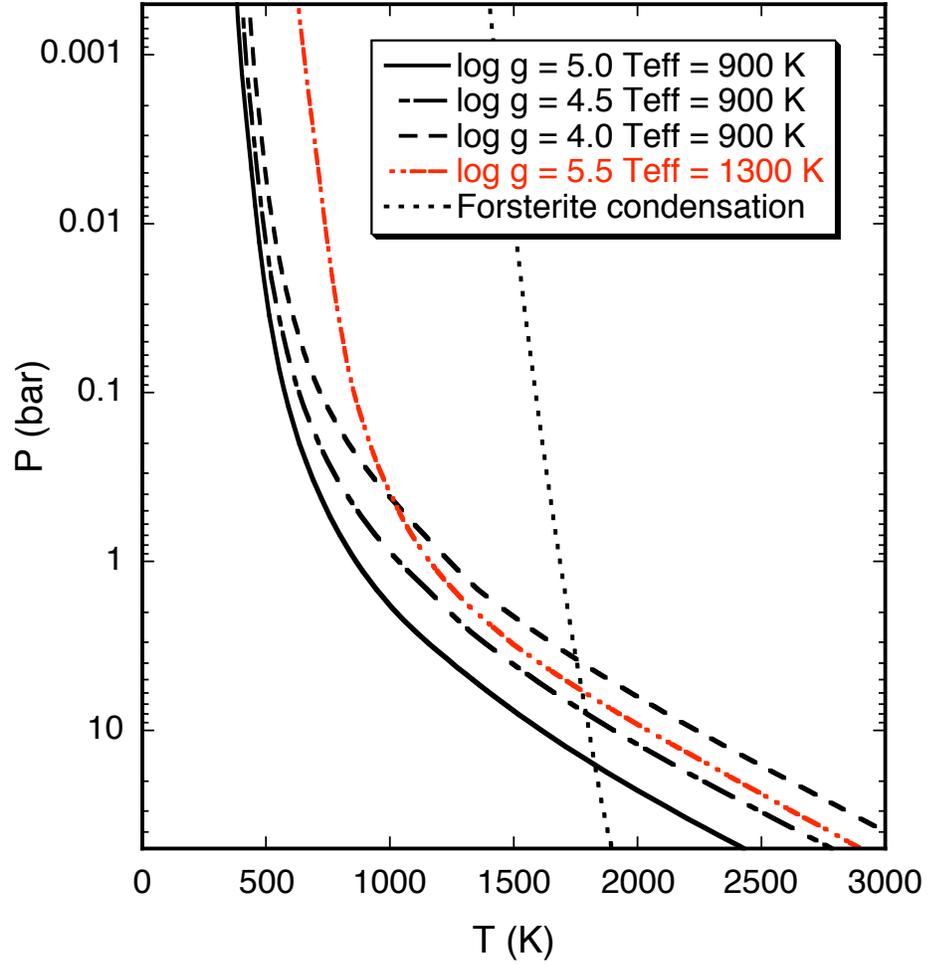}
\caption{Model atmosphere temperature-pressure profiles for cloudy brown dwarfs and planets assuming $f_{\rm sed} = 2$ \citep{Ack01}.  Each profile is labeled with $\log g$ and $T_{\rm eff}$ of the model.   The condensation curve for forsterite is shown with a dotted line.}
\label{profiles}
\end{figure}

\clearpage
\begin{figure}
\includegraphics[angle=0,scale=.15]{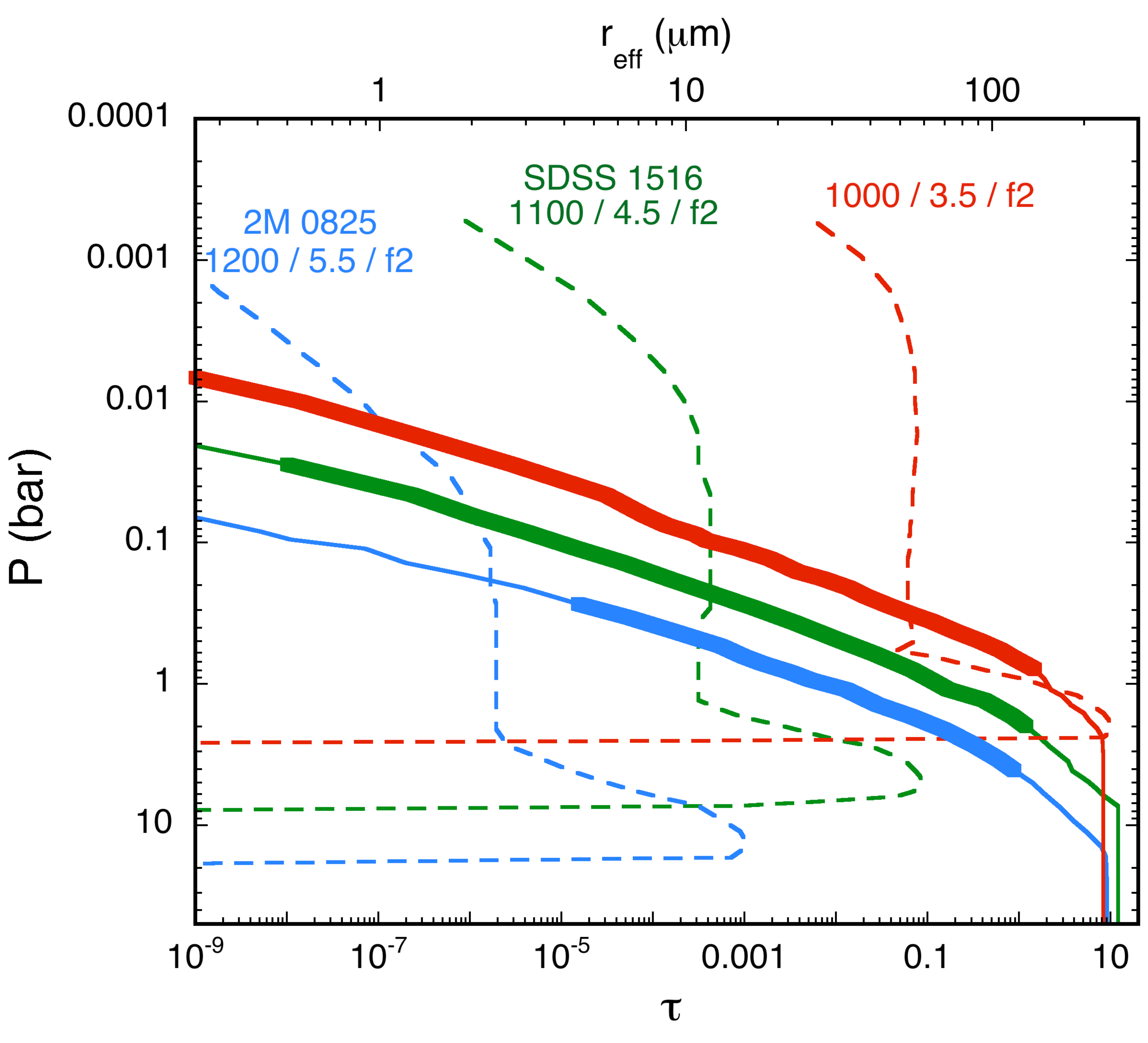}
\caption{Silicate cloud properties as computed by the \citet{Ack01} cloud model for three models.  From left to right the the best-fitting models \cite{Ste09} for 2MASS 0825+21 and SDSS 1516+30 are shown along with a profile for a young, cloudy, three Jupiter mass planet.
Labels underneath each object name denote model $T_{\rm eff} (\rm K )$ / $\log g\,({\rm cgs})$ / $f_{\rm sed}$.  Dashed curves show the effective radius, $r_{\rm eff}$ of the particles on the top axis. The column optical depth as measured from the top of the atmosphere  is shown by the solid lines and the scale on the bottom axis.  Thicker lines denote the region of the cloud which lies within the $\lambda = 1$ to $6\,\rm \mu m$ photosphere.  Other modeled clouds are not shown for clarity. }

\label{fsed}
\end{figure}

\clearpage
\begin{figure}
\includegraphics[angle=0,scale=.75]{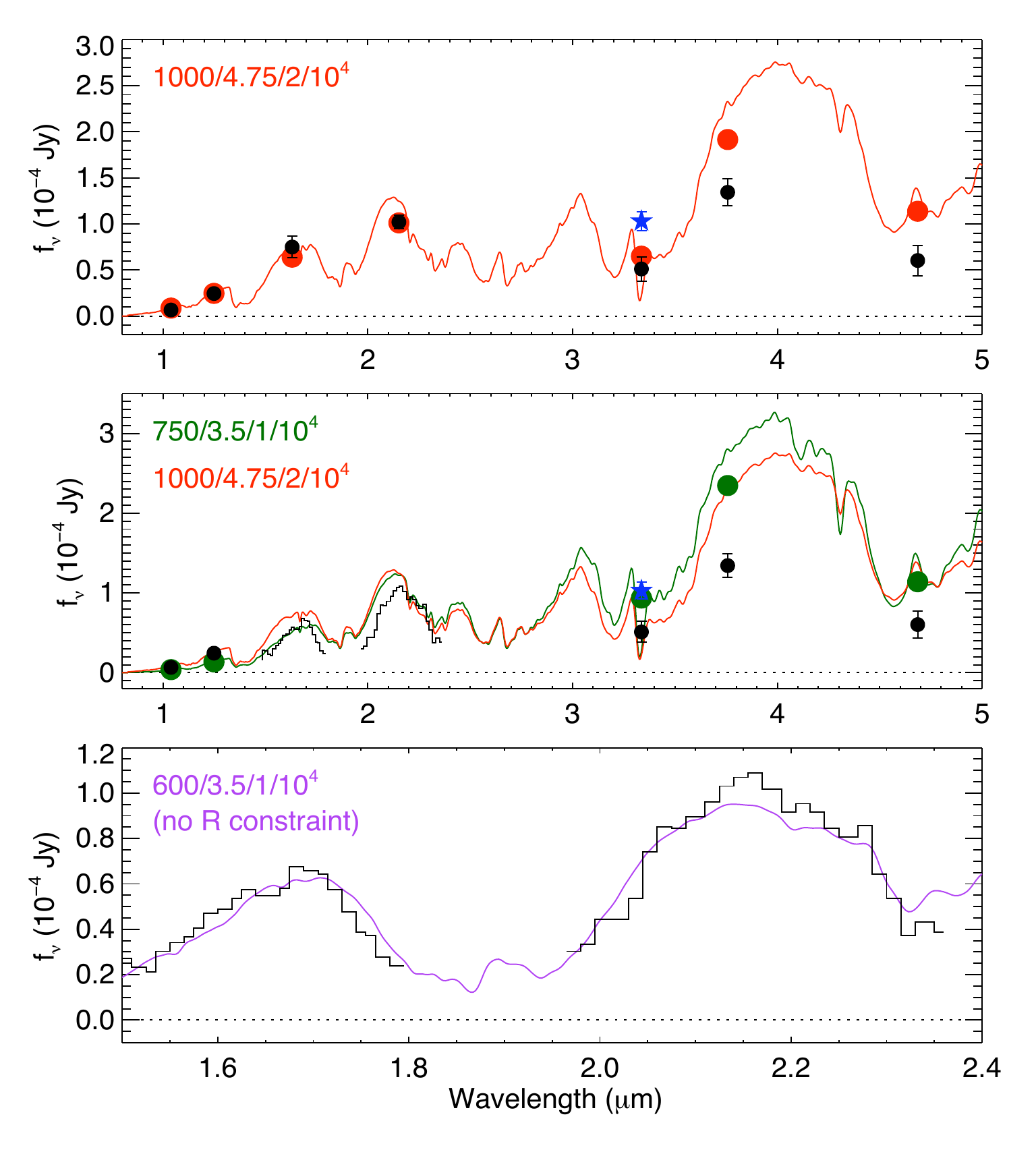}

\caption{Observed (black) and model (red, green, purple) photometry and spectra (see Table 1 and Barman et al. (2011a)) for HR 8799b.  Models are identified in the upper left hand corner of each panel by $T_{\rm eff}/\log g\,({\rm cgs}) /f_{\rm sed}/ K_{zz}$.  The top panel shows the model that best fits the photometry alone while the middle panel shows the solution that best fits both the photometry (excluding $H$ and $K$ bands) and spectroscopy simultaneously.  Model fluxes and photometry have been computed for radii specific to the $T_{\rm eff}$ and $\log g$ of the atmosphere model at a distance of 39.4 pc as observed from Earth.  The [3.3] $\rm \mu m$ photometry of \citet{Ske12} is shown as a blue star and is not included in the fits but rather is shown for comparison purposes only.
The lower panel shows the model that best fits the $H$ and $K$-band spectrum alone.  However in contrast to the top two panels where the absolute flux level of the models are set by the model radii and known distance to HR 8799, the absolute flux level of the model in the lower panel is determined  by minimizing $\chi^2$ between the models and data. }

\label{planet_b}
\end{figure}

\clearpage
\begin{figure}
\includegraphics[angle=0,scale=.9]{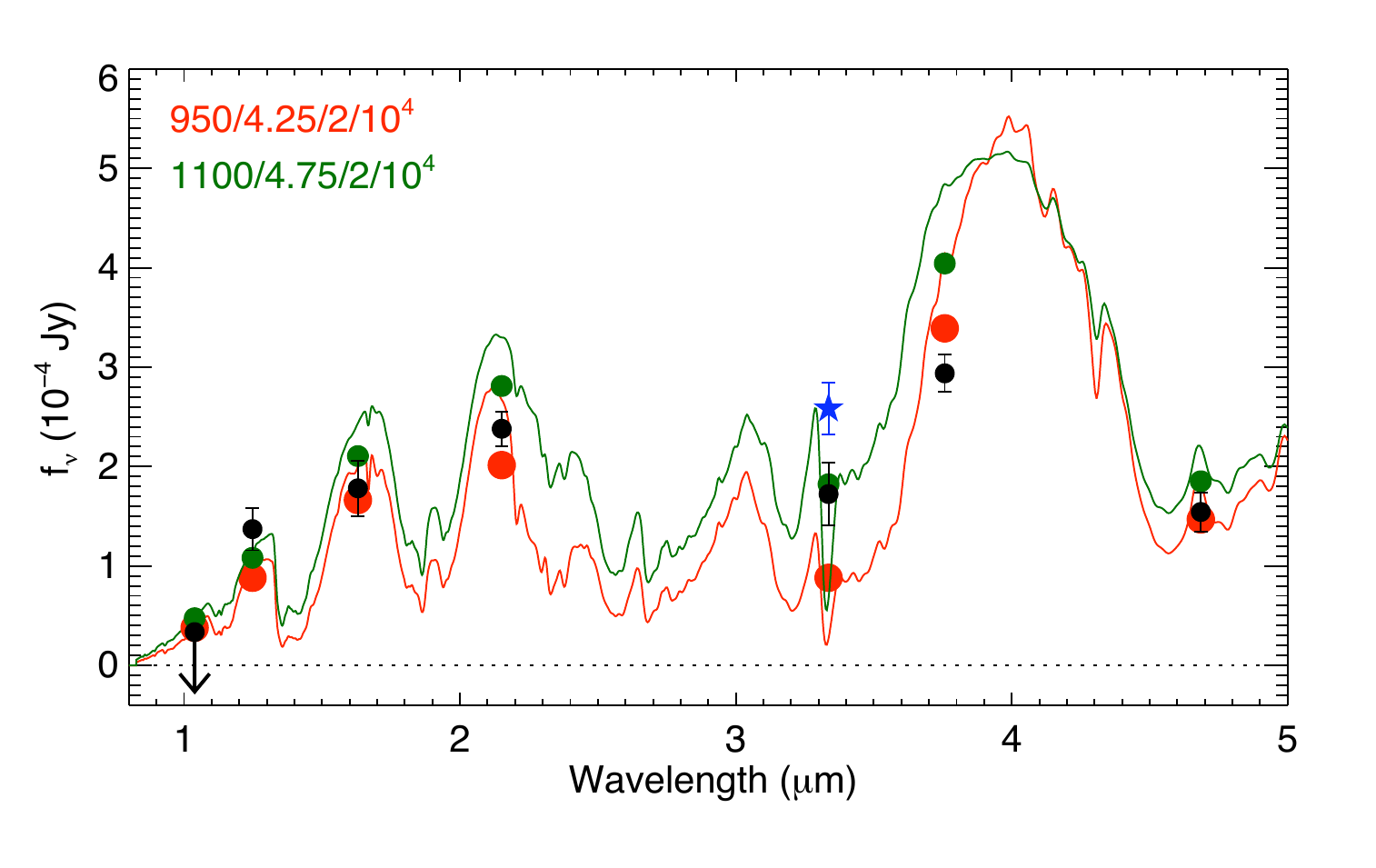}
\caption{The two best fitting model spectra for HR 8799 c. Observed photometry (see Table 2) is shown in black, high and low gravity solutions in green and red, respectively. The two solutions correspond to the centers of the two best fitting islands in the contour plot shown in the middle panel of Figure \ref{contour}. Models are identified in the upper left hand corner by $T_{\rm eff}/\log g\,({\rm cgs}) /f_{\rm sed}/ K_{zz}$. The [3.3] $\rm \mu m$ photometry of \citet{Ske12} is shown as a blue star and is not included in the fits but rather is shown for comparison purposes only. }

\label{planet_c}
\end{figure}

\clearpage
\begin{figure}
\includegraphics[angle=0,scale=.9]{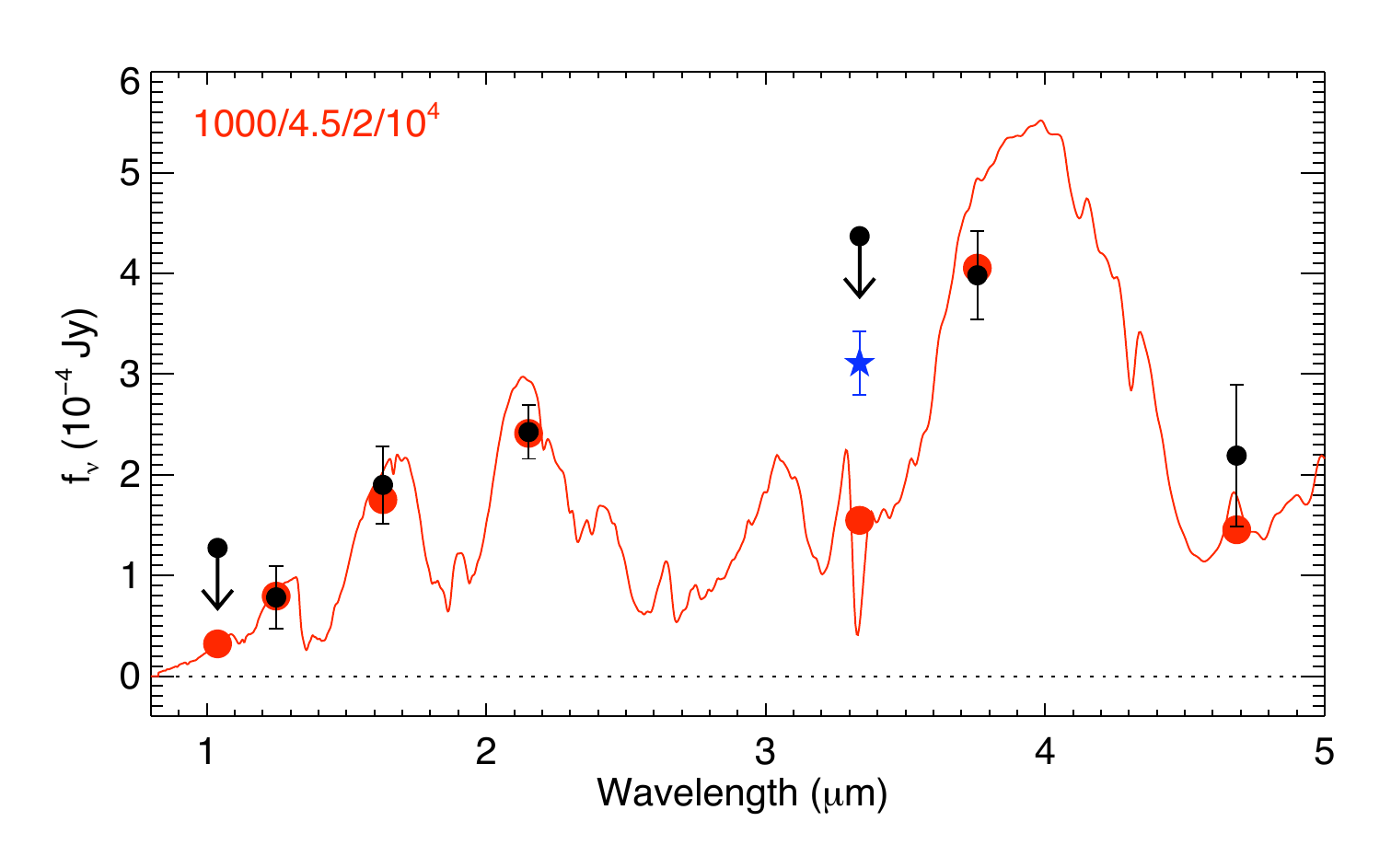}
\caption{The best fitting model for HR 8799 d.  Observed photometry (see Table 1) is shown in black; model photometry is indicated by the red dots. Model is identified in the upper left hand corner by $T_{\rm eff}/\log g\,({\rm cgs}) /f_{\rm sed}/ K_{zz}$. The 3.3-$\rm \mu m$ photometry of \citet{Ske12} is shown as a blue star and is not included in the fits but rather is shown for comparison purposes only. }

\label{planet_d}
\end{figure}

\clearpage
\begin{figure}
\includegraphics[angle=0,scale=.75]{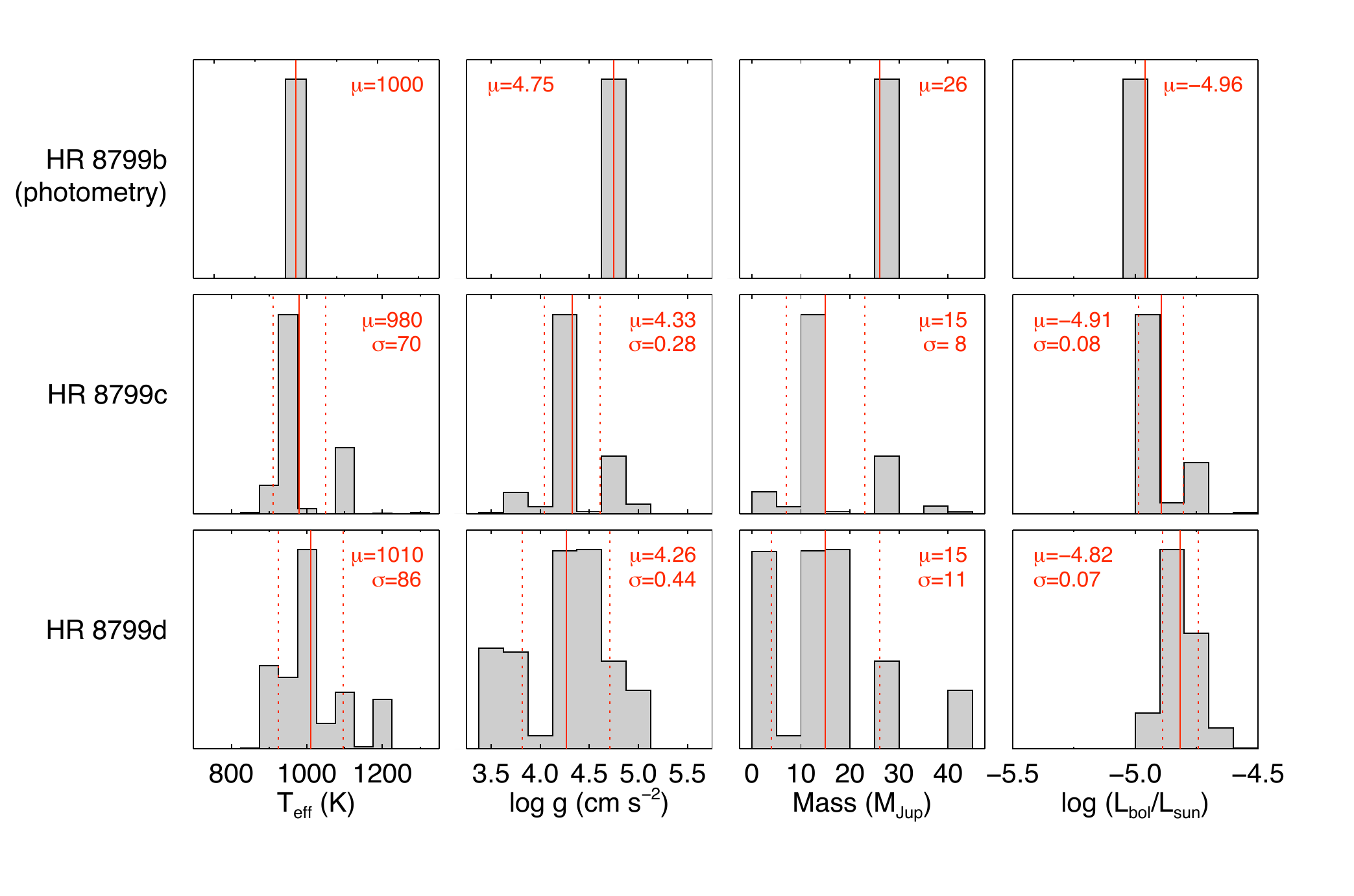}
\caption{Histograms depicting the probability density distributions of the various
model parameters to planets HR 8799 b, c, and d.  For planet b only the results for the fitting of the photometry are shown.  The $T_{\rm eff}$ and $\log g$ histograms can be thought of as the projection of the contours shown in Figure \ref{contour} onto these two orthogonal axes.  In each case the mean of the fit and the standard deviation are indicated by $\mu$ and $\sigma$, respectively.  These quantities are in turn illustrated by the solid and dashed vertical lines. For the parameters for planet b, only a single model is identified so no standard deviation is given. The third and fourth columns of histograms depict the same information as the first two, but for the mass and luminosity corresponding to each $(T_{\rm eff}, \log g)$ pair, as computed by the evolution model. }

\label{hist}
\end{figure}

\clearpage
\begin{figure}
\includegraphics[angle=0,scale=.4]{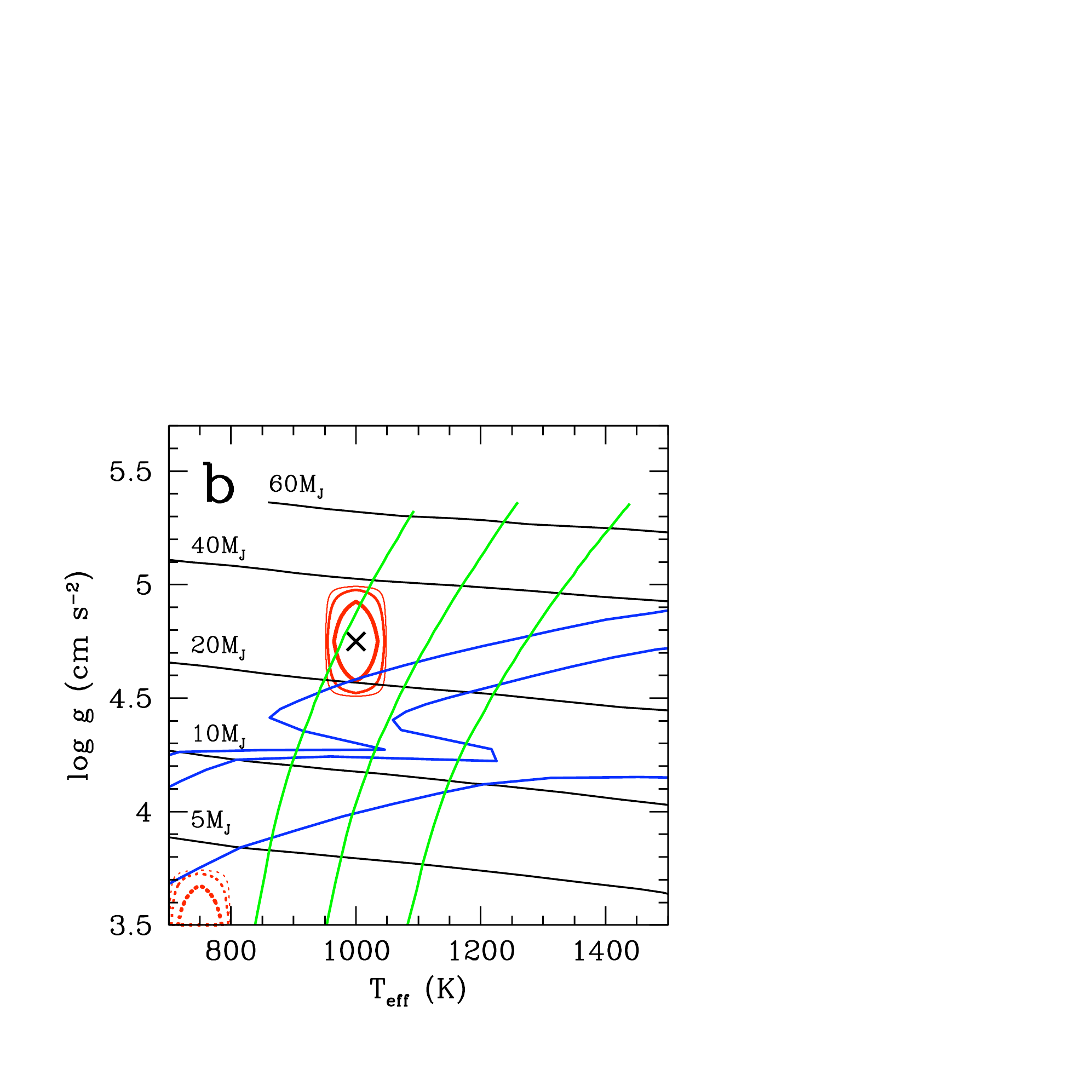}
\includegraphics[angle=0,scale=.4]{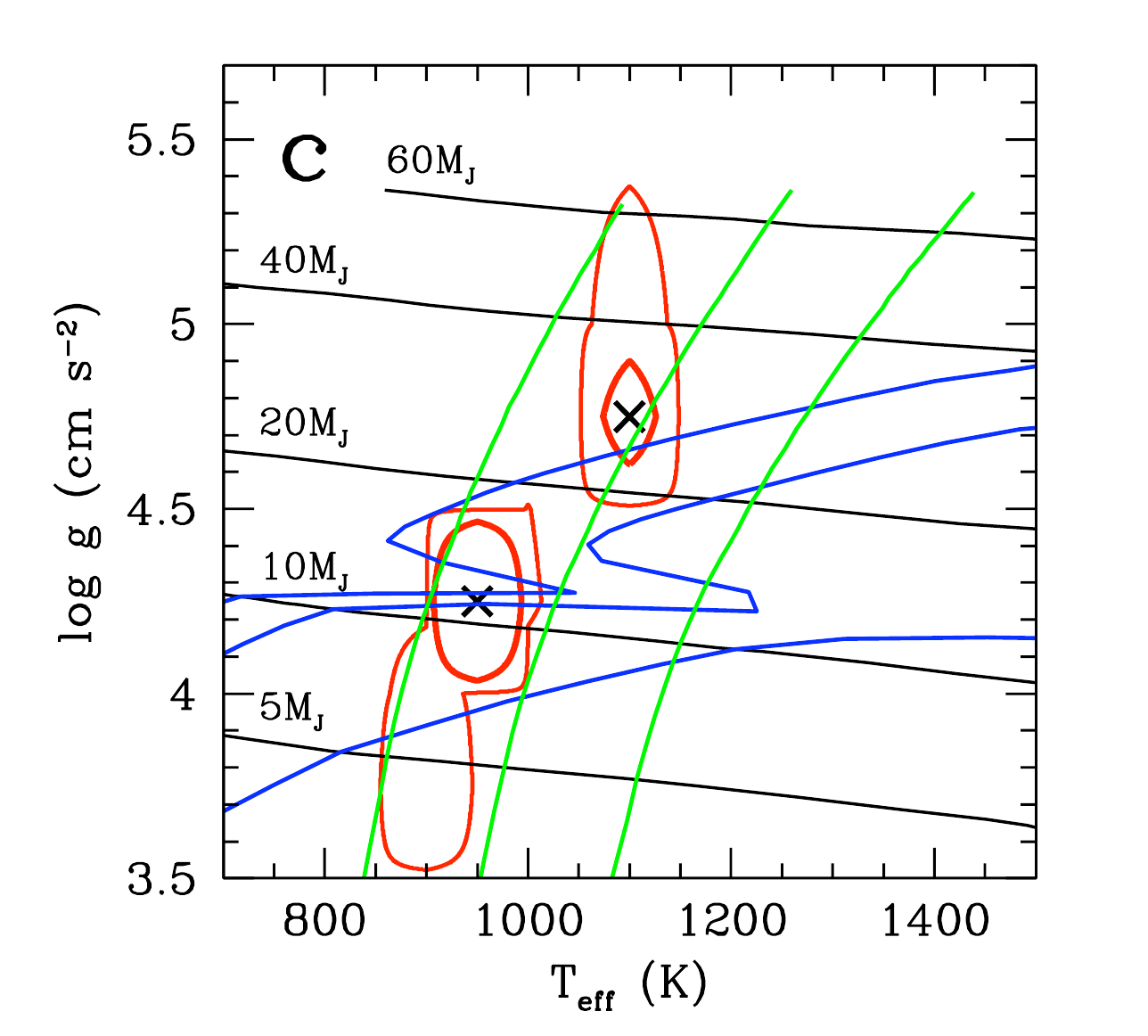}
\includegraphics[angle=0,scale=.4]{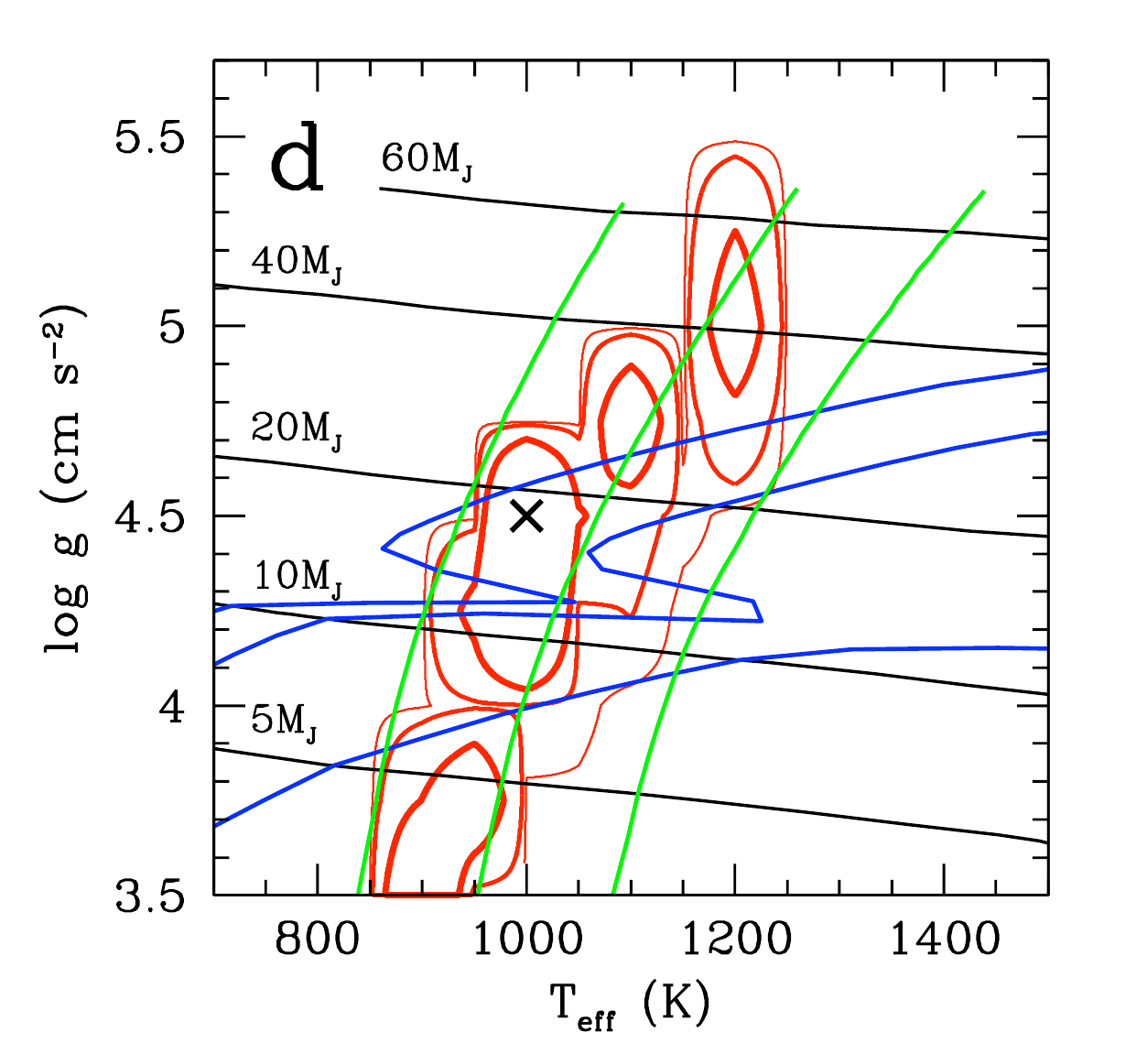}
\caption{Contours illustrate domain of best-fitting models on the $\log g - T_{\rm eff}$ plane. For each planet three contours are shown 
which correspond to integrated probabilities of 68, 95, and 99\%  (red, thick to thin contours). Evolution tracks from \citet{Sau07} are shown as labeled black curves; planets evolve from right to left with time across the diagram as they cool and contract.  Blue curves are isochrones at (bottom to top) 30, 160, and 300 Myr; kinks in the older two isochrones arise from deuterium burning (objects burning D are substantially hotter than lower mass objects of the same age).  Green curves are constant luminosity curves at (left to right) $\log L/L_\odot=-5, -4.75, -4.5$. For planet b solid contours denote fits to only the photometry while dashed curves are fits to photometry and H and K-band spectra. Crosses denote the individual model cases plotted in Figures 4 -- 6. }

\label{contour}
\end{figure}

\clearpage
\begin{figure}
\includegraphics[angle=0,scale=1.0]{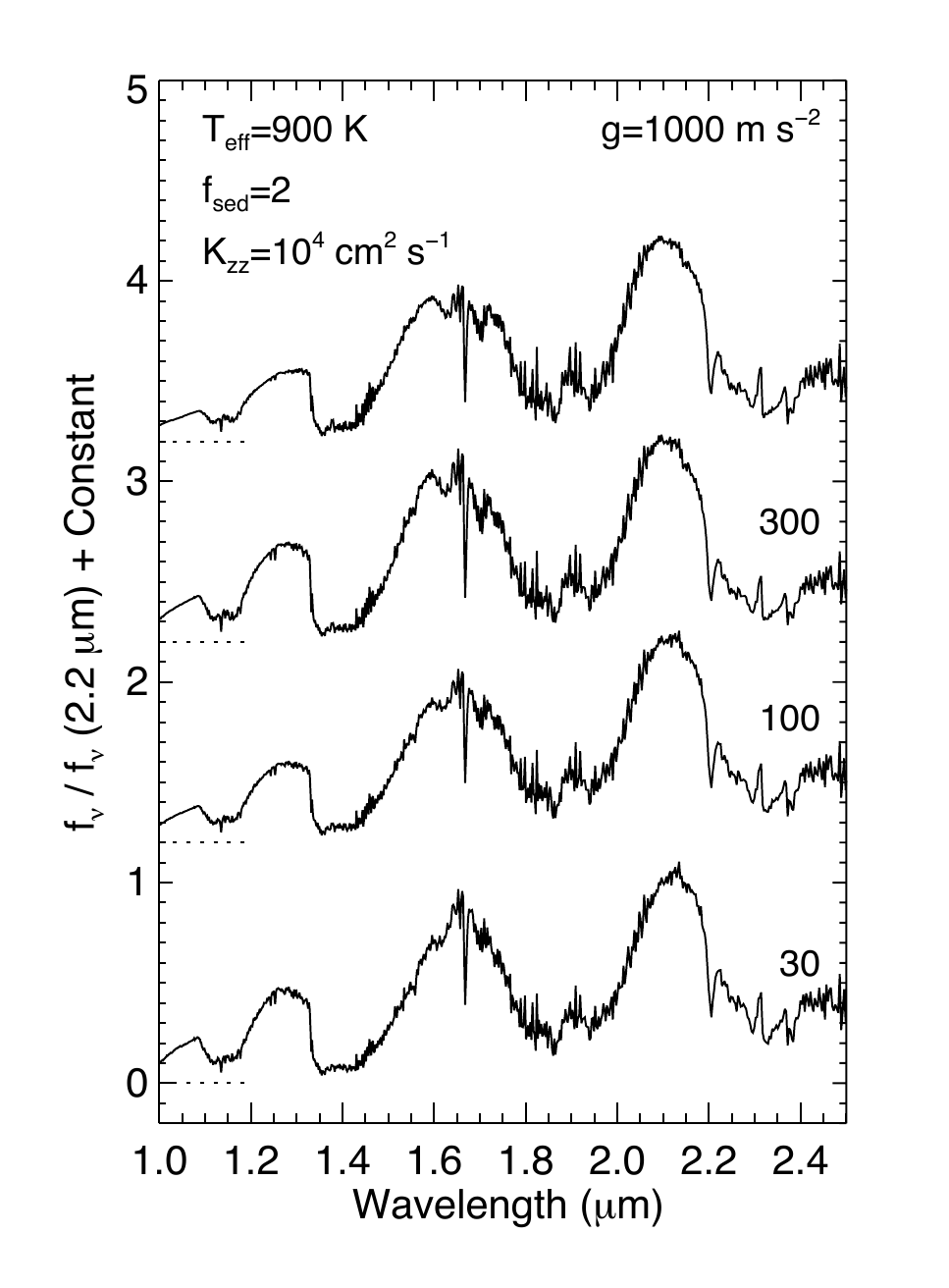}

\caption{Model spectra at fixed $T_{\rm eff} = 900\,\rm K$ and varying gravities (labeled along right hand side), including several of the cases shown in Figure 2.  Models are shown at a spectral resolution $R=1000$. }

\label{models}
\end{figure}

\clearpage

\begin{figure}[ht!]
\begin{center}
 \includegraphics[width=4in]{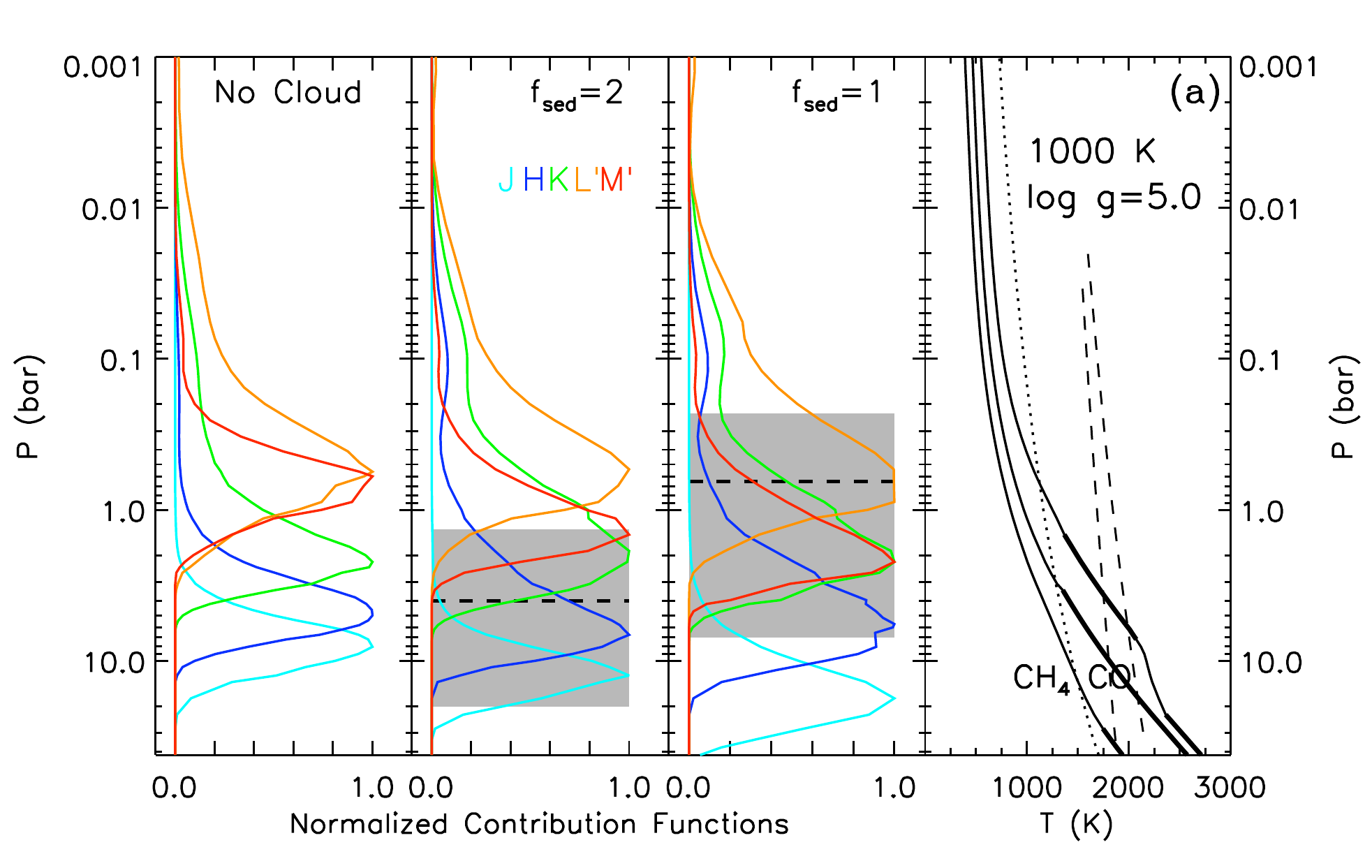}
 \includegraphics[width=4in]{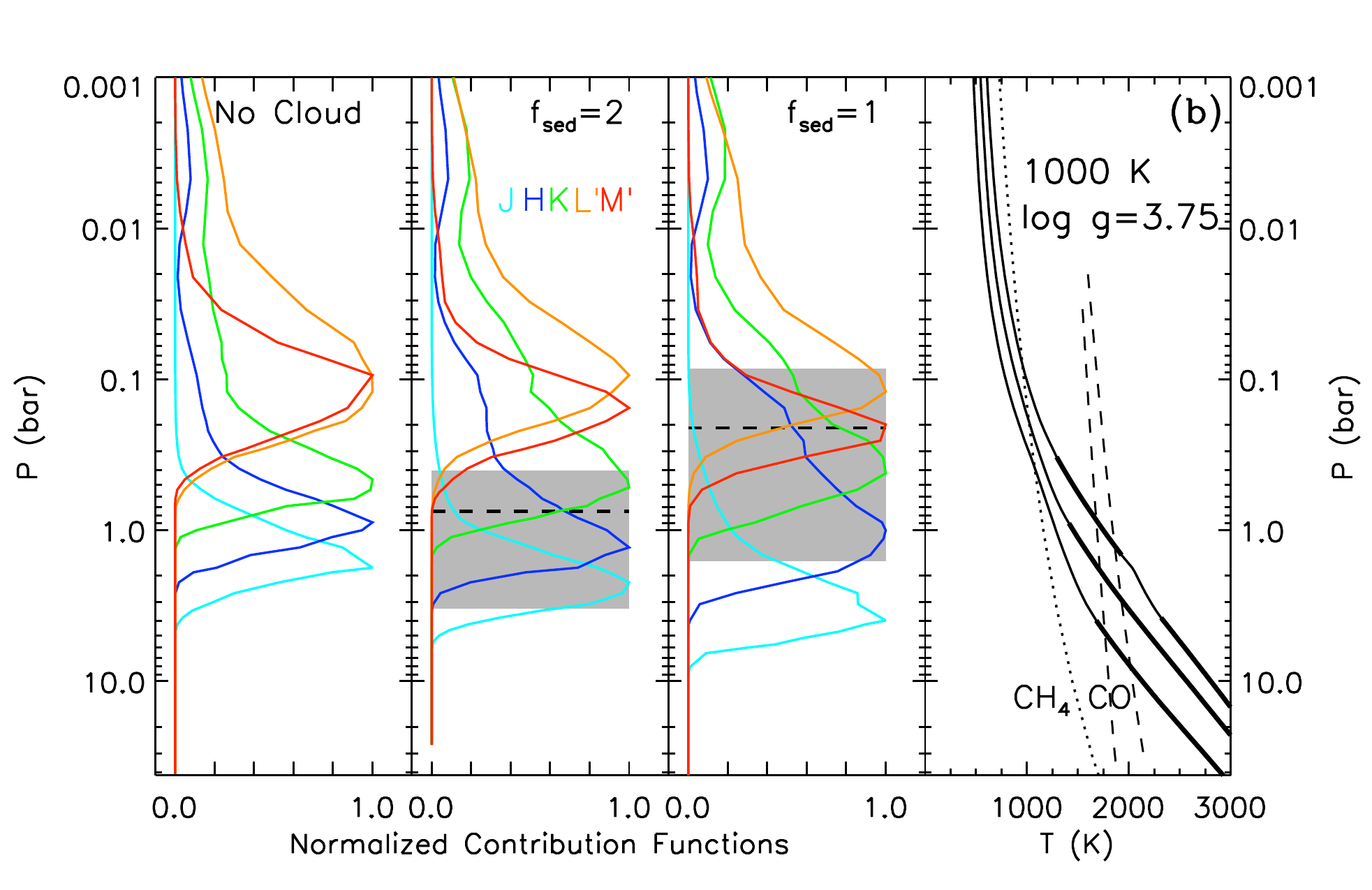}
 \caption{Illustration of the effect of gravity and cloud properties on modeled emergent flux for $T_{\rm eff}=1000\,\rm K$ and $\log g=5.0$ {\bf (a)} and 3.75 {\bf (b)}.  Both plots (a) and (b) consist of four sub-panels.  The right-most sub-panel depicts the $T(P)$ profiles for three atmosphere models with the indicated $T_{\rm eff}$ and $\log g$.  In both cases the profiles are for (left to right) for cloudless, $f_{\rm sed}=2$, and 1 models. Thick lines denote the convective regions of the atmosphere models.  The dotted line denotes chemical equilibrium between CO and $\rm CH_4$.  The dashed lines are the condensation curves for Fe (right) and $\rm Mg_2SiO_4$ (left).  The cloud base is expected at the point where the condensation curves cross the $T(P)$ profiles.  Remaining panels show the contribution function (see text) averaged over the {\it J, H, K,} $L^\prime$ and $M^\prime$ bandpasses (colored lines) for each of the three model cases.  The shaded regions denote the extent of the cloud, extending from the point where the integrated optical depth from the top of the model is 0.1 to the cloud base.  Thick horizontal dashed line denotes cloud $\tau = 2/3$.}
 \label{contrib}
\end{center}
\end{figure}

\clearpage
\begin{figure}
   \plotone{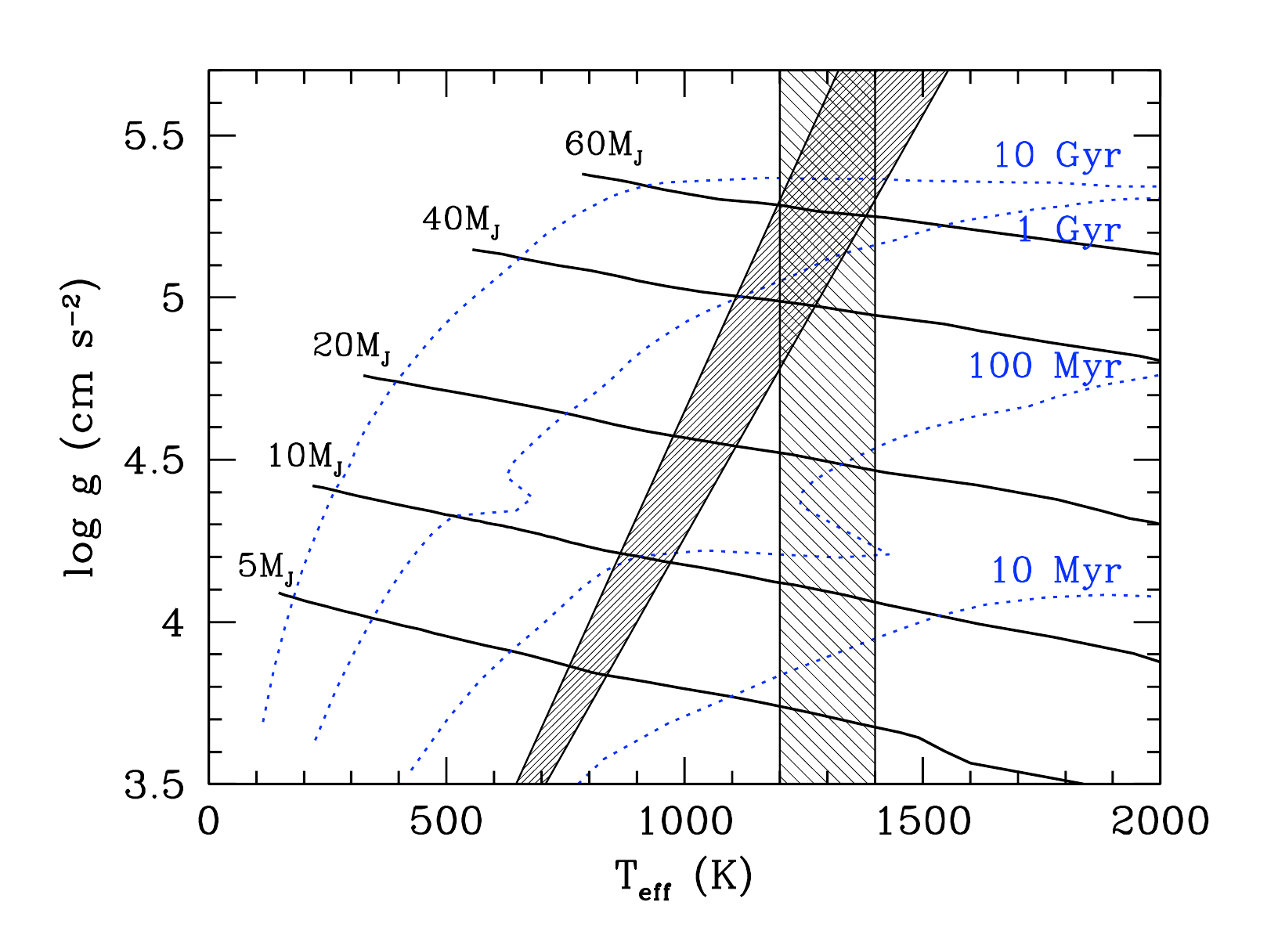}
   \caption{Definition of the transition from cloudy to cloudless surface boundary
            condition for the evolution.  This represents a toy model of the L/T transition.
            In the hybrid toy model of \citet{Sau08}, the transition region is
            independent of gravity and the cloud clearing occurred between $\teff=1400$ and 1200$\,$K
            (lightly hashed area).
            To the right of the transition region shown, the surface boundary condition is based on 
            cloudy atmosphere models; to the left, on cloudless atmospheres; and on a simple interpolation
            in the transition region.  Here, we present an evolution calculation where the
            $\teff$ range of the transition is made gravity dependent (densely hashed area).
            Representative cooling tracks are shown in black and labeled by the mass.  
            Isochrones are the blue dotted lines.}
   \label{trans_gdep}
\end{figure}
\clearpage
\begin{figure}
   \includegraphics[width=5in]{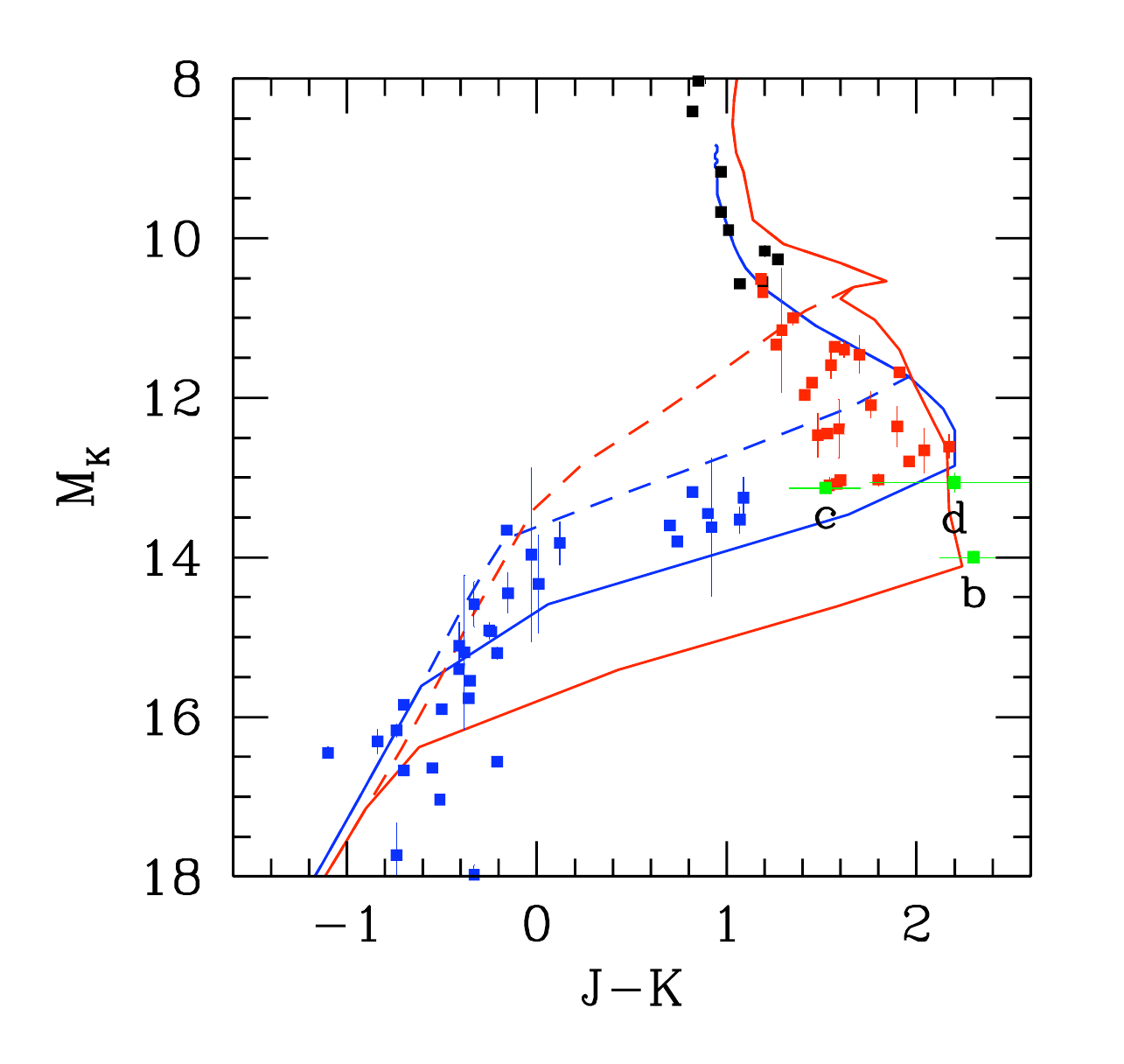}
   \caption{Examples of cooling tracks for objects of 5$\,$M$_{\rm J}$ (red) and 20$\,$M$_{\rm J}$ (blue)
            in a $M_K$ vs. $J-K$ (MKO system) color-magnitude diagram where the transition from cloudy 
            ($\fsed=1$) to cloudless atmospheres is taken into account explicitly as in
            \citet{Sau08}.  Dashed lines show the evolution when the transition occurs over a fixed range of
            $\teff$ that is independent of gravity, solid lines show the evolution for the gravity-dependent 
            transition (see Fig. \ref{trans_gdep}). The planets in the HR 8799 planets are shown with green
            symbols while resolved field objects are shown in black (M dwarfs), red (L dwarfs) and blue (T dwarfs).
            The photometry is from \citet{leggett02}, \citet{Kna04}, \citet{marocco10}
            \citet{mccaugh04}, \citet{burgasser06}, and \citet{ll05}.  The parallaxes are from \citet{perryman97}, 
            \citet{dahn02}, \citet{tbk03}, \citet{Vrb04}, \citet{marocco10}, and various references in 
            \citet{leggett02}.}
   \label{evol_gdep}
\end{figure}

\end{document}